\documentclass[12pt]{article}

\usepackage{amsmath}
\usepackage{setspace}
\usepackage{array}
\usepackage{times}
\usepackage{bm}
\usepackage{graphicx}
\usepackage{latexsym}
\usepackage{color}
\usepackage[font=small,skip=-10pt]{caption}

\numberwithin{figure}{section}

\def\Parrow{\stackrel{P}{\rightarrow}}     
\def\darrow{\stackrel{d}{\rightarrow}}

\usepackage[dvips, letterpaper, nohead, top=1.3in, bottom=1.3in, left=1.3in, right=1.3in]{geometry}

\def\qed{\rule{2mm}{2mm}}

\newtheorem{theorem}{Theorem}[section]
\newtheorem{lemma}{Lemma}[section]

\newtheorem{corollary}{Corollary}[section]
\newtheorem{remark}{Remark}[section]

\normalsize

\begin{document}

\small\normalsize
\title
{ Analysis of error control  in large scale two-stage multiple hypothesis testing}

\author{
Wenge Guo \\
Department of Mathematical Sciences\\
New Jersey Institute of Technology \\
Newark, NJ 07102-1982 \\
\and
Joseph P. Romano \\
Departments of Statistics and Economics\\
 Stanford University\\
Stanford, CA 94305-4065\\
}

\date
{February 25, 2017}

\maketitle
\begin{abstract}
When dealing with the problem of simultaneously testing a large number
of null hypotheses, a natural testing strategy is to first reduce the number of tested hypotheses by some selection (screening or filtering) process, and then to simultaneously test the selected hypotheses. The main advantage of this strategy is to greatly reduce the severe effect of high dimensions. However, the first screening or selection stage must be properly accounted for in order to maintain some type of error control. In this paper, we will introduce a selection rule based on a selection statistic that is independent of the test statistic when the tested hypothesis is true. Combining this selection rule and the conventional Bonferroni procedure, we can develop a powerful and valid two-stage procedure. The introduced procedure has several nice properties: (i) it completely removes the selection effect; (ii) it reduces the multiplicity effect; (iii) it  does not ``waste" data while carrying out both selection and testing. Asymptotic power analysis and simulation studies illustrate that this proposed method can provide higher power compared to usual multiple testing methods while controlling the Type 1 error rate. Optimal selection thresholds are also derived based on our asymptotic analysis.
\end{abstract}



\newpage

\normalsize

\section{Introduction}

Consider the multiple testing problem of simultaneously testing a large number $m$ of hypotheses.
When $m$ is large,
standard multiple testing procedures suffer from low ``power" and are unable to distinguish between null and alternative effects because
extremely small $p$-values are required if one properly accounts for Type 1 error control, such as the familywise error rate (FWER); see Lehmann and Romano (2005).
It is only by weakening the measure of error control,  such as the false discovery rate (FDR), that some discoveries may be found (Benjamin and Hochberg, 1995).  But, such discoveries are not as forceful as when they arise while controlling the FWER.

 When ``most" null hypotheses are ``true", a  common and useful approach is to first reduce the number of hypotheses being testing in order to construct methods which are better
able to distinguish alternative hypotheses.
That is, one applies some selection, filtering or screening technique  based on some selection statistics  in order to reduce the number of hypotheses being tested. Then,  one can use standard stepwise methods to test the reduced number of tests. Such two-stage methods have been extensively used in practice to deal with various problems of multiple testing (McClintick  and Edenberg, 2006; Talloen et al., 2007; Hackstadt and Hess, 2009).
As in the bulk of this paper, such approaches are called two-stage procedures. In the first stage, some screening or selection method is applied in order to reduce
the number of tests.  In the second stage, the reduced number of tests is tested.
 A major limitation of these methods is
there lacks a systematic  consideration of the selection effect.
In other words, one cannot simply apply some method to the reduced number of hypotheses without accounting for selection in error control.
That is, one cannot in general ``forget" about the screening stage. In other words,
 in order to properly control Type 1 error rates, one must in general account for the screening stage
 by considering the error rate conditional on the method of selection.
Otherwise, lose of Type 1 error control, whether it is FDR, FWER, or an alternative measure, results.

 But, if  screening statistics at the first stage are chosen to be independent of the testing statistics at
 the second  stage (at least under the null hypothesis), then error control simplifies
 as the conditional distributions and unconditional distributions of the test statistics
 are the same (at least under its respective null distribution).
Indeed, Bourgon, Gentleman, and Huber (2010) introduced such  a novel approach of independence filtering to avoid the effect of selection, in which the selection or filtering statistics at the first stage  are chosen to be independent of the test statistics (at least when the corresponding  null hypotheses are true). Two new two-stage methods, which respectively combine the approach of independence filtering with the conventional Bonferroni and Benjamini-Hochberg procedures (Benjamini and Hochberg, 1995), are proposed and  shown to control both
the FWER and FDR under independence of test statistics.  By using the same idea of  independence filtering, Dai et al. (2012) develop several two-stage testing procedures to detect gene-environment interaction in genome-wide association studies. Kim and Schliekelman (2016) further discuss some key questions on how to best apply the approach of independence filtering and quantify the effects of the quality of the filter information, the filter cutoff and other factors on the effectiveness of the filter.


Another commonly used approach to avoid the selection effect is sample splitting in which the data is split  in two independent parts.  One uses the first part of the data to construct the selection or filtering statistics and the second part to
construct the test statistics.
By combining sample splitting with conventional stepwise procedures, one can develop two-stage procedures that guarantee  control of Type 1 error rates (Cox, 1975; Rubin, Dudoit, and van der Laan, 2006; Wasserman and Roeder, 2009). These methods completely remove the effect of selection; however, they often result in power loss due to reduced sample size for testing (Skol, et al., 2006; Fithian, Sun and Taylor, 2014).

In recent  years, there has been a growing interest in selective inference (Benjamin and Yekutieli, 2005; Benjamini, 2010; Taylor and Tibshirani, 2015) and several novel breakthroughs have been made in the context of high-dimensional regression (Berk et al 2013; Barber and Cand$\acute{e}$s 2015; Lee et al 2016; Fithian et al. 2014).
All of these developments take model selection rules as given and develop methods to preform valid inference after taking into account selection effects. Along these lines, a number of selective inference/post selection inference methods have been developed for various model selection algorithms
(Barber and Candes, 2016; Benjamini and Bogomolov, 2014; Fithian et al., 2015; Heller et al., 2016;
Tian and Taylor, 2015a, b; Weinstein, Fithian and Benjamini, 2013; Yekutieli; 2012). In this literature, the problem of how to choose selection rules is often overlooked; however, in practice one can often choose a desired selection rule to lead to favorable conditional properties of inference after selection.
In contrast, rather than treat the selected hypotheses as given, we can propose a rule
in both stages so that the overall procedure has good unconditional error control properties.

Another popular way of exploiting  information in the data is, rather than completely eliminating tests under consideration,   to  construct weights for the null hypotheses and then develop data-driven weighted multiple testing procedures (Roeder and Wasserman, 2009; Poisson et al, 2012). The data-driven weighted methods are pretty general and filtering methods can regarded as its special case. A limitation of such methods is that it is not clear how to assign weights in a data-driven way to ensure control of the FWER or FDR. Very recently, by using   ``covariates"
to construct weights which are independent of the test statistics under the null hypotheses, several Bonferroni-based and Benjamini-Hochberg based data driven weighted methods have been developed that increase power while controlling the FWER and FDR, respectively
(Fino and Salmaso, 2007; Ignatiadis, et al, 2016; Li and Barber, 2016; Lei and Fithian, 2016; Ignatiadis and Huber, 2017).
In addition, when developing more powerful multiple testing methods, there are several other ways of using such additional   covariate information that have recently introduced in the literature, such as local FDR based approaches (Cai and Sun, 2009), stratified Benjamini-Hochberg (Yoo, et al., 2010), grouped Benjamini-Hochberg (Hu, Zhao and Zhou, 2010), and single-index modulated method (Du and Zhang, 2014), etc.

In summary, there is a growing literature of approaches to dimension reduction in high dimensional (single and multiple)  hypothesis testing,
including some useful, novel, and somewhat ad hoc procedures.  The contribution of this paper is to perform a detailed error analysis
in a large scale setting.  We consider an ideal Gaussian model, as is often assumed in the literature. as described in the setup in Section 2.
There, we introduce  a specific two-stage procedure that we will analyze and compare later with other procedures.  Control of the FWER
is presented, though the less formal argument  already appears in Bourgon, Gentleman and Huber (2010).  (The analysis applies to the joint but single testing problem of testing all means zero against the alternative that at least one is not, but the exposition emphasizes the multiple testing problem.) The remainder of the paper is new.
In Section 3, under  a large $m$ asymptotic framework with a sparsity assumption on the number of false hypotheses, we present  detection boundaries for mean levels that can (or cannot be) detected by the two-stage procedure.
In Section 4,  a refinement is obtained so that the exact cutoff is calculated.
Section 5 considers the unknown variance case, where the basic finite sample control of the FWER is replaced by asymptotic control,  but the same power analysis holds as when the variance is known.
In Section 6,  we allow for dependence between the test statistics.
Section 7 theoretically compares the two-stage approach with other methods: Bonferroni and split-sample methods.  By proper choice of how to split, the split sample technique can only perform as well as Bonferroni, with neither approach performing as well as the two-stage method.
A simulation study is presented in Section 8.    Both global tests of a single hypothesis (in a high dimensional setting) as well as multiple tests
are considered.  In the former case, the Higher Criticism (Donoho and Jin, 2004; Donoho and Jin, 2015) is also compared (but it cannot readily be used in the multiple testing case).
In both cases, the two-stage approach offers both control of the Type 1 error rate as well as it performs quite well under various scenarios.
In particular, the two-stage method shows good performance even when variances are unequal and especially under dependence.

\section{The setup}

A very stylized Gaussian setup is assumed, as is conventional in large scale testing.
The problem is testing $m$ means from independent populations, where $m$ is large.

Assume that, for $i = 1, \ldots, m$,
a sample of size $n_i$ from a normal population with
unknown mean $\mu_i$ and variance $\sigma_i^2$ is observed; that is,   data $$X_{ij} \stackrel{i.i.d}{\sim} N(\mu_i,\sigma_i^2),~~ ; j=1, \ldots, n_i,$$ where $m$ is the number of hypotheses  of interest representing
the number of samples or populations,  and $n_i$ is the sample size for the $i$th sample.  The $m$ samples
are assumed mutually independent.
When $m$ is large, it is typically assumed that the $\sigma_i$ are known as well, in which case one can take $n_i = 1$ (by sufficiency).  For now, we will assume $n_i = n$ and $\sigma_i = 1$,
though we will  discuss the unknown variances case later.

 For  $i=1 , \ldots , m$, consider testing hypotheses
 $$H_i: \mu_i = 0~~~~vs.~~~~H'_i: \mu_i \neq 0~.$$
 (One may also treat the case of one-sided alternatives with easy modifications.)
 Define the following two statistics
\begin{equation}\label{S}
S_{n,i} = \sum_{j=1}^{n}  X_{i,j}^2
\end{equation}
and
\begin{equation}\label{equation:T}
T_{n,i} = \frac{\sqrt{n}\overline{X}_{n,i}}{\hat \sigma_{n,i}},
\end{equation}
where $\overline{X}_{n,i}$ and $\hat \sigma_{n,i}^2$ are respectively the sample mean and (unbiased) sample variance for the $i$th sample, i.e.,
$\overline{X}_{n,i} = \frac{1}{n}\sum_{j=1}^{n}  X_{i,j}$ and $\hat \sigma_{n,i}^2 = \frac{1}{n-1}\sum_{j=1}^{n} (X_{i,j} - \overline{X}_{n,i})^2$.

The basic two-stage strategy for our method is as follows.  The statistics $S_{n,i}$ are first used
to ``select" which of the hypotheses to ``test" in the second stage, at which point the  statistics
$T_{n,i}$ are used.
There are various choices for the selection statistics, as well as test statistics.   For example, one could
use the $t$-statistics $T_{n,i}$ in both stages.  Regardless, the first consideration would
then be how to set critical values in each stage in order to ensure some measure of  Type 1 error control, such as the familywise error rate (FWER), the probability of at least one false rejection.   We will be specific about the critical values
soon, but the key motivation for the choice of the sum of squares selection statistic $S_{n,i}$
and test statistic $T_{n, i}$ is based on the following well-known facts.
First, under $H_i:
 \mu_i = 0$ (and $\sigma_i = 1$) we have that $$
 S_{n,i}   \sim \chi_{n}^2~~~{\rm  and~~~}
 T_{n,i}  \sim t_{n-1}~;$$
 that is, $S_{n,i}$ has the Chi-squared distribution with $n$ degrees
 of freedom and $T_{n,i}$ has the $t$-distribution with $n-1$ degrees of freedom.
 But, the more important reason motivating our choice is that,  by Basu's theorem, $S_{n,i}$ and $T_{n,i}$ are independent under $H_i$ (Lehmann and Romano, 2005).
 Note that $E ( S_{n,i} ) = n  +  n \mu_i^2$, so that larger values of $S_{n,i}$ are indicative of larger
 values of $\mu_i^2$.

A simple selection rule is used for selecting which  hypotheses $H_i$ are to be tested
at the second stage.   Given a threshold $u$, $H_i$ is selected iff $S_{n,i}  \ge u$.
Let $\hat S_n$ denote the indices of selected hypotheses,
with $| \hat S_n |$ the number of selected hypotheses.
At the second stage, one can simply apply the Bonferroni test; that is,
reject $H_i$ iff $|T_{n,i} | \ge t_{n-1}(1 - \frac{\alpha}{2 | \hat S_n |})$, the $1 - \alpha/2 | \hat S_n |$
quantile of the $t$-distribution with $n-1$ degrees of freedom.

\begin{lemma}\label{lemma:control} For any choice  of the  threshold $u$, the above two-stage procedure
controls the FWER at level $\alpha$.
\end{lemma}

Like all proofs, see the appendix.

\begin{remark}\rm  The proof of Lemma \ref{lemma:control} requires that any test statistic $T_{n,i}$ be independent of the
selection statistics $S_{n,1} , \ldots , S_{n,m}$, if $H_i$ is true.  Note that it is not required that the test statistics $T_{n,1} , \ldots , T_{n,m} $
are jointly independent of the selection statistics.

More generally, the two-stage procedure controls the familywise error rate when any test statistic is independent of the selection statistics,
even outside our stylized Gaussian model.
\end{remark}

The simple two-stage method can be improved by a Holm-type stepdown improvement.   To describe the method, simply apply the Holm method (Holm, 1979)
to the $p$-values based on the selected set of hypotheses.
More specifically,  let $\hat p_{n,i}$ denote the marginal $p$-value when testing $H_i$ based on $T_{n,i}$.  Of course, in the model above,
this is just the probability that a $t$-distribution with $n-1$ degrees of freedom exceeds the observed value of $| T_{n,i} |$.
Let $\tilde p_{n,i}$ be one if $H_i$ is not selected and equal to $\hat p_{n,i}$ if it is selected.
Let
$$\tilde p_{n, r_1} \le \tilde p_{n, r_2} \le \cdots \le \tilde p_{n, r_m}$$
denote the ordered $p$-values, so that $r_i$ is the index of the $i$th most significant $p$-value.
Now, apply Holm's procedure based on the $p$-values $\tilde p_{n,r_i}$ with $ 1 \le i \le | \hat S_n| $.
Thus, $H_{r_i}$ is rejected if $\tilde p_{n, r_j} \le \alpha / (  |  \hat  S_n| - j + 1 )$ for $j  = 1 , \ldots , i$.

\begin{theorem}\label{theorem:holm}
Under the setting of Lemma \ref{lemma:control}, apply the Holm method to the selected set of hypotheses. Then, this modified procedure
controls the FWER at level $\alpha$.
\end{theorem}

Thus, one can do even better by using a Holm-like  stepdown method, or even
a stepdown version of Sidak's procedure; see Lehmann and Romano (2005) and Guo and Romano (2007).
Indeed, conditional on the selection statistics, all computed  true null  $p$-values based on ``detection" statistics  at the second stage
are conditionally uniform on $(0,1)$ and hence unconditionally as well.
Thus, any  multiple testing method based on $p$-values is available.
For example, one can also apply the Benjamini-Hochberg procedure based on the selected
$p$-values for controlling the false discovery rate (Benjamini and Hochberg, 1995).  In all such cases, the motivation is that gains are possible because
at the second stage only a reduced number of hypotheses are tested, with the hopes
of increased ability to detect or discover false null hypotheses.
Furthermore,  both the selection and detection stages are based on the full data
(rather than a split sample approach which is used to obtain independence of
the stages) and there is no selection effect because of independence between the selection and test statistics when the corresponding hypothesis is true.

So far, the threshold for selection has been just generically  set at some constant $u$. We
now discuss this choice.
For our method, we will choose $u$ of the form  $u  = \chi_n^2(1-\beta)$, the $1-\beta$ quantile of $\chi_n^2$. Since $S_{n,i} \sim \chi_n^2$ when $H_i$ is true, such a selection threshold $\chi_n^2(1-\beta)$  ensures that roughly $\beta m$ hypotheses are selected, at least
if most null hypotheses are true.  The question now is how to choose $\beta$.
Let $m' = m^{\gamma}$ and $\beta = \frac{m'}{m} = m^{-(1-\gamma)}$, where $\gamma$ is a given positive constant satisfying $0 < \gamma \le 1$.
Then, roughly $\beta m = m^{\gamma} = m'$ hypotheses
are selected for testing.  A choice of $\gamma$ must still be specified.

Since Type 1 error control is ensured regardless of the choice of $\gamma$, we now
turn to studying the power of the procedure.
In our asymptotic analysis, the following is assumed.

\medskip

\textbf{Assumption A:} $\lim_{m \rightarrow \infty} \frac{\log m}{n} = d, 0 \le d < \infty$, where $d$ is a nonnegative constant.

\medskip

Note that as $m$ is equal to $10,000$, $100,000$ or $1,000,000$, the values of $\log m$ are respectively $9, 12$ and $14$. So, it is reasonable and  often sufficient to characterize the relationship between $m$ and $n$ by imposing  Assumption A.
In applications, $ m$ and $n$ (and hence $\log(m) / n$) are known, and generally we will have $0 \le d \le 1$.
We will consider the probability of rejecting a null hypothesis $H_j$
having mean $\mu_j \ne 0 $, which  without loss of generality  can be taken to be positive.
Further assume without loss of generality that it is  $H_1$ that is false with mean $\mu_1 > 0$.
If $\mu_1$ is constant, then under Assumptions A and $d > 0$ , we have $\sqrt{n}\mu_1 = O(\sqrt{2 \log m})$.
On the other hand, if $\mu_1$ varies with $m$ (and $n$) such that  $\mu_1(m) \rightarrow \infty$ as $m$ approaches  infinity, then $\lim_{m \rightarrow \infty} \frac{\sqrt{n}\mu_1}{\sqrt{2 \log m}} = \infty.$ Finally, if the sample size $n$ is very large,  so that $\log (m) $ is very small compared to the sample size $n$, then the value of $d$ should be taken to be 0.
In the following, we mainly perform   asymptotic power analyses under Assumption A.
Sometimes, $d > 0$ is assumed, in which case the case  $d = 0$  can
either be treated separately with ease, or by a limiting argument as $d$ tends to zero.

\section{Power analysis of two-stage procedure}

In order to analyze the power of the  two-stage procedure, we break up the analysis in two parts.
The first part analyzes the probability of  ``selection" in the first stage, while the second
will analyze the probability of ``detection" in the second stage.
Rejection of $H_i$ then occurs when both $H_i$ has been selected at the first stage
and then detection occurs at the second stage.
Roughly, the basic goal will be to determine how large in absolute value an alternative mean must be in order to ensure that  the probability of rejection tends to one.

\subsection{The probability of selecting $\mu_1$}

Consider the case where $\mu_1 > 0$ is a constant, so that $H_1$ is false.   We now consider the asymptotic behavior
of the probability that $H_1$ is selected in the first stage of the two-stage procedure.
Recall that
 $\chi_n^2(1-\beta)$ denotes the $1- \beta$ quantile of $\chi_n^2$,  the Chi-squared distribution
 with $n$ degrees of freedom, i.e.,
$P \left( \chi_n^2 \ge \chi_n^2(1-\beta) \right) = \beta.$
Hypothesis $H_1$ is selected if $S_{n,1} > \chi_n^2 ( 1 - m^{\gamma -1} )$.

\begin{lemma}\label{lemma:select}

\noindent (i) Under Assumption A,  if
\begin{equation}\label{equation:w1}
\mu_1^2 > 2(1 - \gamma)d + 2 \sqrt{(1 - \gamma)d}~,
\end{equation}
then
 $$\lim_{m \rightarrow \infty} P_{\mu_1} \{H_1 \text{ selected} \} = 1~.$$

\noindent (ii)   Under Assumption A,  if
\begin{equation}\label{equation:w2}
\mu_1^2 < 2(1 - \gamma)d + \frac{1}{4} \sqrt{(1 - \gamma)d}~,
\end{equation} then
$$\lim_{m \rightarrow \infty} P_{\mu_1} \{H_1 \text{ selected} \} = 0~.$$

\end{lemma}

In Lemma \ref{lemma:select}(i), if $d = 0$, then the condition (\ref{equation:w1}) always holds, while in (ii) if $d = 0$
the condition (\ref{equation:w2}) never holds, which implies $H_1$ is selected with probability tending to one.

Note that there exists a  gap between the two detection thresholds in Lemma \ref{lemma:select}, but we will derive an improved, exact result in  Section 4.

\subsection{The probability of detecting $\mu_1$}

We now consider the probability that $\mu_1$ is detected at the second stage using the $t$-statistic
$T_{n,1}$.  That is, we now analyze  the probability that $| T_{n,1}|$ exceeds
$t_{n-1} ( 1 - \frac{\alpha}{2 | \hat S_n | } )$, regardless of whether or not $H_1$ is selected
at the first stage.  Later, we will analyze the two stages jointly, but  for now note that if $H_1$ is false,
then it is no longer the case that the selection statistic $S_{n,1}$ and the detection statistic
$T_{n,1}$ are independent.

First, in order to understand the detection probability, we need to understand $| \hat S_n |$, the number of selections  from the first stage (as it is random).
 Let $I_{m,0}$ denote the indices of true null hypotheses from $1$ to $m$, and let
  $I_{m,1}$ denote the indices of false null hypotheses from $1$ to $m$.
  Let $| I_{m,0} |$ and $| I_{m,1} |$ denote the number of true and false null hypotheses, respectively, from $1, \ldots , m$.

 We will assume some degree of sparsity in the sense
 \begin{equation}\label{equation:sparse} | I_{m,1} | \asymp m^{1- \epsilon }
 \end{equation}
  for some $0 < \epsilon \le 1$.  We will even allow $\epsilon = 1$, treating
  the ``needle in the haystack" problem,
  where exactly  one alternative hypothesis is true.

\begin{lemma}\label{lemma:cheb}
The number of selected hypotheses $| \hat S_n |$ satisfies
\begin{equation}\label{equation:selecta}
E ( | \hat S_n | ) \ge m^{\gamma} \to \infty~.
\end{equation}
If we assume the sparsity condition (\ref{equation:sparse}), then
 \begin{equation}\label{equation:cheb}
|  \hat S_n |  / m^{\gamma} \Parrow 1~,
 \end{equation}
 and
\begin{equation}\label{equation:selectb}
| \hat S_n | / E (|  \hat S_n |) \Parrow 1~.
\end{equation}
 as long as $\epsilon + \gamma > 1$.

 \end{lemma}

\begin{lemma}\label{lemma:detect}  Under Assumptions A  and (\ref{equation:sparse}), we have
\begin{itemize}
  \item[(i)] when $\mu_1^2 > e^{2\gamma d} - 1$, $\lim_{m \rightarrow \infty} P_{\mu_1} \{H_1 \text{ detected}\} = 1$;
  \item[(ii)] when $\mu_1^2 < e^{2\gamma d} - 1$, $\lim_{m \rightarrow \infty} P_{\mu_1} \{H_1 \text{ detected}\} = 0$.
\end{itemize}
\end{lemma}

Obviously, if $d = 0$, then $P_{\mu_1}  \{ H_1~ {\rm rejected} \} \to 1$ for any $\mu_1 > 0$.

\subsection{Asymptotic power analysis}

We now combine the two stages to determine the value of $\mu_i$ that leads to rejection of $H_i$.
Let $A_i$ be the event that $H_i$ is selected in the first stage and let $B_i$
be the event that $|T_{n,i} | > t_{n-1} ( 1- \frac{\alpha}{2 | \hat S_n | } )$ at the second stage.
Note $A_i$ and $B_i$ are dependent in general.
Then, the power of the two-stage method, i.e., the probability that $H_i$ is rejected, is
\begin{equation}
\text{Power}  = P_{\mu_1}  \{ A_i \bigcap B_i \} = P_{\mu_1} \{ A_i \} - P_{\mu_1} \{ A_i \bigcap B_i^c \}
\ge P_{\mu_1}
\{ A_i  \} - P \{ B_i^c \}~.
\end{equation}
Therefore, in order for rejection of $H_i$ to occur with probability tending to one, it
is sufficient to show both $A_i$ and $B_i$ have probability tending to one.
Also, we have
\begin{equation}
\text{Power} \le  \min \{  P_{\mu_1} \{ A_i \} , P_{\mu_1} \{ B_i \} \}~.
\end{equation}

 Combining Lemma \ref{lemma:detect} and \ref{lemma:select}, the following result holds.

\begin{theorem}\label{theorem:31}
Under Assumption A and (\ref{equation:sparse}), we have
\begin{itemize}
  \item[(i)] when $\mu_1^2 > \max\{e^{2\gamma d} - 1, 2(1 - \gamma)d + 2 \sqrt{(1 - \gamma)d} \}$, $$\lim_{m \rightarrow \infty} P_{\mu_1} \{H_1 \text{  rejected}\} = 1~;$$

  \item[(ii)] when $\mu_1^2 < \max\{e^{2\gamma d} - 1, 2(1 - \gamma)d + \frac{1}{4} \sqrt{(1 - \gamma)d} \}~,$ $$\lim_{m \rightarrow \infty} P_{\mu_1} \{H_1 \text{ rejected}\} = 0~.$$
\end{itemize}
\end{theorem}

\begin{corollary}\label{corollary:31}
 Under Assumption A  with $d = 0$, for any given $0 < \gamma \le 1$,  (\ref{equation:sparse}) and any $\mu_1 \neq 0$,
$$\lim_{m \rightarrow \infty} \text{Pr}_{\mu_1} \{H_1 \text{ rejected}\} = 1.$$
\end{corollary}

Of course, in multiple testing problems, there are many notions of power one might wish to maximize: the probability of rejecting at least one false null hypothesis,
the probability of rejecting all false null hypotheses,
the probability of rejecting at least $k$ false null hypotheses (for any given $k$),
the expected number (or proportion)  of rejections among false null hypothesis,  etc.
Theorem \ref{theorem:31} and Corollary \ref{corollary:31} apply directly  to the expected proportion of false null hypotheses rejected.
For example,  in the setting where all false null hypotheses have a common mean $\mu_1$,  then the expected proportion of correct rejections
equals the probability that any one of them is rejected, which tends to one (or not) based on the threshold for $\mu_1$.

\section{Further improvement}

In order to improve Theorem \ref{theorem:31}, we need to derive improved bounds on extreme Chi-squared quantiles.
(Note the slack in the bounds provided in Lemmas \ref{lemma:LM} and \ref{lemma:I}.)

Let
\begin{equation}\label{equation:gx}
g(x) = \frac{e^x - 1 -x}{x^2}~,
\end{equation} which is increasing on $(0, \infty)$. Then, define
\begin{equation}\label{equation:ac}
a(c) = \left[g^{-1}\left(2/c^2\right)/c\right]^2~,
\end{equation}
which is decreasing in $c$.

\begin{lemma}\label{lemma:cstar}
Given the value $\gamma$ used in stage one  for selection with
$\beta_m  = m^{\gamma -1}$, and $d$ in Assumption A, with $d > 0$,
define $c^* = c^* ( \gamma , d )$ to be the solution of the equation
\begin{equation}\label{equation:cstar}
a (c) = ( 1- \gamma ) d~.
\end{equation}

\noindent (i)   For any $c > c^*$ and sufficiently large $n$,
\begin{equation}\label{equation:cstarle}
\chi_n^2(1-\beta_m) \le n + 2\log\left(\frac{1}{\beta_m }\right) + c\sqrt{n \log\left(\frac{1}{\beta_m }\right)}.
\end{equation}

\noindent (i)   For any $c  <  c^*$ and sufficiently large $n$,
\begin{equation}\label{equation:cstarge}
\chi_n^2(1-\beta_m ) \ge n + 2\log\left(\frac{1}{\beta_m }\right) + c\sqrt{n \log\left(\frac{1}{\beta_m }\right)}.
\end{equation}

\end{lemma}

 Based on Lemma \ref{lemma:cstar},  Lemma \ref{lemma:select} can be improved as follows.

\begin{lemma}\label{lemma:improve}
\noindent Under Assumption A and (\ref{equation:sparse}), we have
\begin{itemize}
  \item[(i)] when
$\mu_1^2 > 2(1 - \gamma)d + c^*(\gamma, d) \sqrt{(1 - \gamma)d}$,  $\lim_{m \rightarrow \infty} P_{\mu_1} \{H_1 \text{ selected} \} = 1$.
  \item[(ii)]  when
$\mu_1^2 < 2(1 - \gamma)d + c^*(\gamma, d) \sqrt{(1 - \gamma)d}$,  $\lim_{m \rightarrow \infty} P_{\mu_1} \{H_1 \text{ selected} \} = 0$.
\end{itemize}
\end{lemma}

Combining Lemma \ref{lemma:improve} and Lemma \ref{lemma:detect}, Theorem \ref{theorem:31} can be improved as follows.

\begin{theorem}\label{theorem:41} Under Assumption A and (\ref{equation:sparse}),  we have
\begin{itemize}
  \item[(i)] when $\mu_1^2 > \max\{e^{2\gamma d} - 1, 2(1 - \gamma)d + c^*(\gamma, d) \sqrt{(1 - \gamma)d} \}$, $$\lim_{m \rightarrow \infty} P_{\mu_1} \{H_1 \text{ rejected}\} = 1~~;$$
  \item[(ii)] when $\mu_1^2 < \max\{e^{2\gamma d} - 1, 2(1 - \gamma)d + c^*(\gamma, d) \sqrt{(1 - \gamma)d} \}$, $$\lim_{m \rightarrow \infty} P_{\mu_1} \{H_1 \text{ rejected}\} = 0~.$$
\end{itemize}
\end{theorem}

\begin{remark}\rm

Theorem \ref{theorem:41} offers an approach of determining the value of tuning parameter $\gamma$. By minimizing the right-hand side of the inequality in Theorem \ref{theorem:41} (i) or (ii) with respect to $\gamma$, one can determine an optimal value $\gamma^*$ of $\gamma$ for each given value of $d$, which maximizes probability of detecting any false null or average power asymptotically. As seen from Figure 4.1 (left), the chosen value $\gamma^*$ of $\gamma$ is decreasing in $d$. Note that $d = \lim_{m \rightarrow \infty} \frac{\log m}{n}$, thus $\gamma^*$ is roughly increasing in $n$ if $m$ is fixed and decreasing in $m$ if $n$ is fixed.
For instance, suppose $m = 20,000$ and $n = 20$, then $d \simeq 0.5$. By checking Figure 4.1 (left), the determined value $\gamma^*$ of $\gamma$ is about $0.7$, which implies that about $m^{\gamma^*} = 1,025$ hypotheses are selected in the first stage for detection.

Based on the optimal value $\gamma^*$ of $\gamma$, we can determine by Theorem \ref{theorem:41} the upper bound
of squared mean $\mu_1^2$ for our suggested two-stage Bonferroni procedure, which constitutes a sharp detection threshold. When $\mu_1^2$ is larger than the bound, we can always detect $\mu_1 $. Similarly, we can also determine by Theorem \ref{theorem:bonpower} the detection threshold of $\mu_1^2$ for the conventional Bonferroni procedure. Figure 4.1 (right) shows the detection thresholds of $\mu_1^2$ for these two procedures. As seen from Figure 4.1 (right), the detection thresholds of our suggested procedure are always lower than those of conventional Bonferroni procedure for different values of $d$, and their differences are increasingly larger with increasing $d$.
This implies that our suggested two-stage Bonferroni procedure is more powerful than the conventional Bonferorni procedure and its power improvement over the Bonferroni procedure becomes increasingly larger with increasing $d$. Specifically, the detection threshold of our suggested procedure is almost linear in terms of $d$ with the slope being about $2.001$ and that of the conventional  Bonferroni procedure is an exponential function of $d$.

\begin{figure}\label{figure:4.1}
\begin{center}
\includegraphics[scale=0.7]{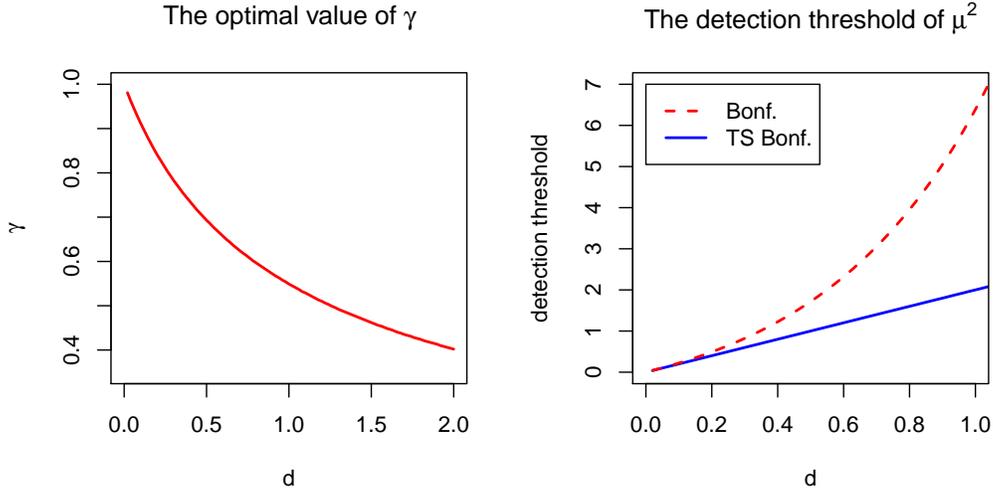}
\end{center}
\caption{The optimal value  (left panel) of the selection parameter $\gamma$ and the corresponding detection threshold (right panel) of squared mean $\mu_1^2$ in Theorem \ref{theorem:41} for our proposed two-stage Bonferroni procedure (TS Bonf.) along with the detection threshold of $\mu_1^2$ in Theorem \ref{theorem:bonpower} for the conventional Bonferroni procedure (Bonf.).}
\end{figure}

\end{remark}

\section{Estimating $\sigma$}

The goal of this section is to show asymptotic control of the FWER is retained when $\sigma_i^2$ are the same as unknown $\sigma^2$ and
$\sigma^2$ is estimated.  To this end,  let $\hat \sigma^2$ denote an overall estimator of $\sigma^2$
which satisfies
\begin{equation}\label{equation:hats}
\hat \sigma^2 - \sigma^2 = O_P \left ( \frac{1}{\sqrt{mn}} \right )~;
\end{equation}
actually, (\ref{equation:hats}) can be weakened but it holds if we take the average or median
of the $m$ sample variances computed from each of the $m$ samples.
Consider the modified procedure based on the selection set
\begin{equation}\label{equation:mselect}
\hat I_n ( u ) = \{ i: S_{n,i} > \hat \sigma^2 u \}~,
\end{equation}
where $u =  \chi_n^2 ( 1- \beta )$ and $\beta = m^{\gamma -1}$ is the critical value used
in selection when it is known that $\sigma = 1$.
The modified two-stage procedure is identical in the second stage in that, for each $i  \in \hat I_n (u)$,
$H_i$ is rejected if its corresponding $t$-statistic  $T_{n,i}$ exceeds the $1 - \alpha / 2 | \hat I_n ( u )| $
quantile of the $t$-distribution with $n-1$ degrees of freedom,
where $| \hat I_n ( u ) |$ denotes the number of selected hypotheses at the first stage.

\begin{theorem}\label{theorem:esigma}  Assume Assumption A.

\noindent (i)   For $\gamma > 1/2$, the above modified two-stage procedure asymptotically controls the
familywise error rate as $m \to \infty$.

\noindent (ii) For $\gamma = 1/2$ and $d > 0$, the above modified
two-stage procedure asymptotically controls the familywise error rate
as $m \to \infty$.
In fact, the same is true if
$$\gamma >  \frac{1}{2} \left [ 1 - \frac{\epsilon^*}{d} + \frac{ \log ( 1 + \epsilon^* )}{d} \right ]~,$$
where $$\epsilon^* = 2 ( 1- \gamma ) d + c^* ( \gamma , d ) \sqrt{ ( 1- \gamma ) d },$$
and $c^* ( \gamma ,d )$ defined in (\ref{equation:cstar}).
\end{theorem}

\begin{remark}\rm
The power analysis used to derive  Theorems \ref{theorem:31} and \ref{theorem:41}  applies equally well to the above  modified  procedure when $\sigma$ is estimated.
Of course, at the second stage, the detection probability analysis remains completely unchanged since there is no modification in the second stage.
In the first stage, the argument for selection can be used along with the assumption (\ref{equation:hats}) to yield the same results, as the argument
is basically the same.
\end{remark}

\section{Dependence}

We now extend the two-stage method when the tests are dependent.  The setup is similar to that described in Section 2.
Assume we have  i.i.d. observations $X_1 , \ldots , X_n$, where $X_j = ( X_{1,j} , \ldots , X_{m,j} )'$ and the $m$ components of $X_j$ may be dependent.
As before, $X_{i,j}$ is $N (  \mu_i , \sigma^2 )$.
(Note that it is not necessary to assume $X_j$ is multivariate Gaussian, but just that the one-dimensional marginal distributions are Gaussian.)
We firstly discuss the case of known $\sigma$. For convenience, we still assume $\sigma = 1$. The two-stage procedure is based on the same selection statistic $S_{n,i}$ and detection statistic $T_{n,i}$ as before.
The two-stage procedure  selects any $H_i$ for which $S_{n,i} > u$ and then rejects $H_i$ if also $|T_{n,i}|$ exceeds $t_{n-1} ( 1- \frac{\alpha}{2 | \hat S_n | } )$,
where $\hat S_n$ is the set of indices $i$ such that $S_{n,i}  > u$ and  $| \hat S_n |$ is the number of selections at the first stage. Let $u = \chi_n^2(1 - m^{\gamma - 1})$ and $\hat S_{n,0}$ be the set of  indices of the selected true null hypotheses, i.e.,
$$\hat S_{n,0} = \{i \in I_{m, 0}: S_{n,i} > u\}.$$
We make the following assumptions regarding $|I_{m, 0}|$ and $| \hat S_{n,0} |$, in which the assumption regarding $| \hat S_{n,0} |$ was already shown to hold under independence in Lemma 3.2.

\medskip

\textbf{Assumption B1:} $~\frac{|I_{m, 0}|}{m} \rightarrow \pi_0~$ as $m \rightarrow \infty,$ where $0 < \pi_0 \le 1$ is a fixed constant.

\noindent In assumption B1, $\pi_0 = 1$ corresponds to sparsity. By assumption B1, we have

\begin{equation}\label{equation:assumB1}
\frac{E\{| \hat S_{n,0} |\}}{m^\gamma } =  \frac{|I_{m, 0}|}{m} \rightarrow \pi_0 \text{ as } m \rightarrow \infty,
\end{equation}
so one can expect the following assumption B2:

\medskip

\textbf{Assumption B2:} $~\frac{| \hat S_{n,0} |}{m^\gamma} \Parrow \pi_0~$ as $m \rightarrow \infty$.

\medskip

\noindent Based on (\ref{equation:assumB1}), to show assumption B2, one just needs
$$\text{Var}\left(\frac{| \hat S_{n,0} |}{m^\gamma}\right) \rightarrow 0,$$
which holds under weak dependence.

\begin{theorem}\label{theorem:dependence1} Assume Assumptions B1 and B2.
The two-stage procedure discussed in Lemma \ref{lemma:control} with $u = \chi_n^2(1 - m^{\gamma - 1})$
asymptotically controls the familywise error rate at level $\alpha$.
\end{theorem}

\begin{remark}\rm
It is interesting to note that in Theorem \ref{theorem:dependence1}, we do not make any assumption of dependence on false null statistics. Only some weak dependence is imposed on true null statistics.
\end{remark}

\begin{remark}\rm
By checking the whole proof of Theorem \ref{theorem:dependence1}, one can see that if the following assumption instead of B2 is imposed,
$$\liminf_{m \rightarrow \infty} \frac{| \hat S_n |}{m^\gamma} \ge 1~,$$
Theorem \ref{theorem:dependence1} still holds.
\end{remark}

\begin{remark}\label{Remark: block depend}\rm
When the selection statistics $S_{n,i}$ are weakly dependent, assumption B2 is satisfied. In the following, we present such an example of block dependence satisfying assumption B2.

Let $I_i = I(S_{n,i} > u)$ for $i = 1, \ldots, m$. Suppose $(I_i)_{i \in I_{m, 0}}$ forming $g$ blocks of sizes $s$ each, which are reformulated as $(\widetilde{I}_{i,j})_{j=1}^s$ for $i=1, \ldots, g$ blocks, are independent to each other, with $|I_{m, 0}| = g s \le m$, $|I_{m, 0}|/m \rightarrow \pi_0~$ and $s/m^\gamma \rightarrow 0$ as $m \rightarrow \infty$, where $0 < \pi_0 \le 1$. Note that $\text{E}(\widetilde{I}_{i,j}) = m^{-(1-\gamma)}.$ In the following, we show that assumption B2 is satisfied under such block dependence. Note that
$$| \hat S_{n,0} | = \sum_{i \in I_{m, 0}} I_i = \sum_{i = 1}^{g} \sum_{j = 1}^{s} \widetilde{I}_{i,j}~.$$
Thus, by block independence of $\widetilde{I}_{i,j}$, we have
$$\text{Var}(| \hat S_{n,0} |) = \sum_{i = 1}^{g} \text{Var} \left(\sum_{j = 1}^{s} \widetilde{I}_{i,j} \right) \le \sum_{i = 1}^{g} \left (\sum_{j = 1}^{s} \text{Var}^{1/2}(\widetilde{I}_{i,j}) \right )^2~.$$
We know that
$$\text{Var}(\widetilde{I}_{i,j}) = \text{E}(\widetilde{I}_{i,j})\left(1 - \text{E}(\widetilde{I}_{i,j}) \right) \le \text{E}(\widetilde{I}_{i,j}) = m^{-(1-\gamma)}~.$$
Combining the above two inequalities,
$$\text{Var}(| \hat S_{n,0} |/m^\gamma) \le gs^2m^{-(1-\gamma)}/m^{2\gamma} \le s/m^\gamma \rightarrow 0 \text{ as } m \rightarrow \infty~.$$
Note that
$$E\left\{| \hat S_{n,0} |/m^\gamma \right\} \rightarrow \pi_0 \text{ as } m \rightarrow \infty~.$$
By Chebychev's inequality, we have
$$| \hat S_{n,0} |/m^\gamma \Parrow \pi_0 \text{ as } m \rightarrow \infty~,$$
and thus assumption B2 is satisfied. ~\qed
\end{remark}

When $\sigma_i^2$ are the same as unknown $\sigma^2$ and $\sigma^2$ is estimated,
we consider the modified two-stage procedure discussed in Theorem \ref{theorem:esigma}. By using similar arguments as in the proof of Theorem \ref{theorem:dependence1}, we can also show that asymptotic control of the FWER  is retained for this procedure under dependence.

For any given $0 < c_n < 1$ and  $u = \chi_n^2(1 - m^{\gamma - 1})$, define
$$\hat S_{n,0}(c_n) = \{i \in I_{m, 0}: S_{n,i} > c_n\sigma^2 u\}.$$ Except for assumption B1, we also make the following two assumptions regarding $\hat \sigma^2$ and $\hat S_{n,0}(c_n)$:

\medskip

\textbf{Assumption B3:} $~\hat \sigma^2 - \sigma^2 = O_P\left(\frac{1}{\sqrt{mn}}\right)$.

\medskip

\textbf{Assumption B4:} $~\frac{| \hat S_{n,0}(1-\delta_n) |}{m^\gamma} \Parrow \pi_0~$ as $m \rightarrow \infty,$ where $\delta_n = \frac{\tau_n}{\sqrt{mn}}$ for some $\tau_n \rightarrow \infty$ slowly.

\medskip

\noindent We should note that assumption B3 has been presented in Section 5 and assumption B4 is a slight extension of assumption B2.

\begin{theorem}\label{theorem:dependence2} Assume Assumptions B1, B3 and B4.
The two-stage procedure discussed in Theorem \ref{theorem:esigma}
asymptotically controls the familywise error rate at level $\alpha$.
\end{theorem}

When the selection statistics $S_{n,i}$ are block dependent, if the overall estimate $\hat \sigma^2$ is chosen as
$$\hat \sigma^2 = \frac{1}{m} \sum_{i=1}^m \hat \sigma^2_{n, i}~,$$
we can similarly show that assumptions B3 and B4 are satisfied under block dependence by using the similar arguments as in the case of known variance where we showed in Remark \ref{Remark: block depend} that assumption B2 is satisfied under block dependence.

\section{Alternative Methods}

In this section, we perform a corresponding power analysis with some alternative methods.

\subsection{Bonferroni}

First, we consider the Bonferroni method, which rejects $H_i$ if $|T_{n,i} | > t_{n-1} ( 1 - \frac{\alpha}{2 m} )$.
We consider the power or rejection probability of $H_i$ when $\mu_i$ is the mean.

\begin{theorem}\label{theorem:bonpower}
Assume Assumption A. For the original Bonferroni method,

\noindent (i)  when $\mu_1^2 > e^{2d} -1$,
$$\lim_{m \to \infty} P_{\mu_1} \{ H_1~{\rm rejected} \} = 1~.$$

\noindent (ii)  when $\mu_1^2 < e^{2d} -1$,
$$\lim_{m \to \infty} P_{\mu_1} \{ H_1~{\rm rejected} \} = 0~.$$
\end{theorem}

\begin{remark} \rm
In Theorem \ref{theorem:bonpower}, if $d = 0$, then the stated condition in (i) always holds,
which implies $H_1$ is rejected by the Bonferroni procedure with probability tending to one. On the other hand, the stated condition in (ii)  holds for any large $\mu$ if $d$  is large enough,
which implies $H_1$ is rejected with probability tending to zero.
\end{remark}

\begin{remark}\rm
In the case of known variance, one can use a $z$-statistic with a normal quantile $z_{1 - \frac{\alpha}{2m}}$.  Similar to the proof of Theorem
 \ref{theorem:bonpower}, it can be shown that the threshold $e^{2d} -1$ can be replaced by $2d$.
\end{remark}

\subsection{Split Sample Method}

A common way (Skol et al. 2006; Wasserman and Roeder 2009) to achieve a reduction in the number of tests is to split the sample in two $n = n_1 + n_2$ independent parts. The first part, based on $n_1$
observations is used to determine which hypotheses will be selected. Then, those selected hypotheses are tested based on the independent set of $n_2$ observations.
Since the two subsamples are independent (as we have been assuming all $n$ observations are i.i.d.),  it is easy to control the FWER.
Indeed, suppose the first subsample produces a reduced set of hypotheses with indices $\hat S_n$, so that the number of selected hypotheses is $|\hat S_n |$.
Then, the Bonferroni procedure applied to the remaining $n_2$ observations evidently controls the FWER.
Specifically,   for $k = 1,2$, suppose $T_{n,i}^{(k)}$ denotes the $t$-statistic computed on the $k$th subsample of size $n_k$ for testing $H_i$.
Here, $H_i$ is selected if $| T_{n,i}^{(1)} | > u$, for some cutoff $u$.  Here, we will take $u$ to be of the form
$$u = t_{n_1 -1} ( 1 - m^{\gamma -1}/2 )$$
for some $0 < \gamma \le 1$.  If $|\hat S_n |$ denotes the number of $T_{n,i}^{(1)}$ satisfying the inequality so that $H_i$ is selected, then
$H_i$ is rejected at the second stage if also
$$|T_{n ,i}^{(2)}| > t_{n_2 -1}  ( 1- \frac{ \alpha}{2 | \hat S_n | })~.$$
For any cutoff $u$ used for selection, this procedure controls the FWER.
We would like to determine the smallest value of $|\mu_1|$  where such a procedure has limiting power one.

\begin{theorem}\label{theorem:split}
Assume Assumption A.  Also assume $n_1 / n \to r$ and the sparsity condition (\ref{equation:sparse}).
For the above split sample method,

\noindent (i)  when $\mu_1^2 >  \max \left [  \exp ( \frac{ 2 ( 1- \gamma ) d}{r} ), \exp ( \frac{ 2 \gamma d}{1-r} )   \right ]-1$,
$$\lim_{m \to \infty} P_{\mu_1} \{ H_1~{\rm rejected} \} = 1~.$$

\noindent (ii)  when $\mu_1^2 <  \max \left [  \exp ( \frac{ 2 ( 1- \gamma ) d}{r} ), \exp ( \frac{ 2 \gamma d}{1-r} )   \right ]-1$,
$$\lim_{m \to \infty} P_{\mu_1} \{ H_1~{\rm rejected} \} = 0~.$$
\end{theorem}

\begin{remark} \rm
By Theorem \ref{theorem:split}, the detection threshold (or rather its square) of the split sample method is equal to
$$\max \left [  \exp ( \frac{ 2 ( 1- \gamma ) d}{r} ), \exp ( \frac{ 2 \gamma d}{1-r} )   \right ]-1~,$$
which depends on $d$, which we set as $\log (m)/n$,  a choice of $\gamma$, as well as the choice of $r$ to determine the split sample sizes.
We want the threshold to be as small as possible.  With $d$ fixed, minimizing over both $\gamma $ and $r$ requires
minimizing $\max [ (1- \gamma )/r , \gamma / (1-r) ]$.
If $r$ is fixed, the optimizing choice of $\gamma $ is $\gamma = 1-r$, in which case
the threshold becomes $\exp (2d) -1$, which is the same as the original Bonferroni procedure.
Note that there are infinitely many optimizing combinations of $r$ and $\gamma$ as long as $\gamma = 1-r$. Regardless, no claim can be made to an improvement over the Bonferroni procedure.
(On the other hand, we could also apply the split sample method and then apply Holm method in the second stage, which if compared
to the usual Holm method based on the full data could offer an improvement because critical values now change more rapidly
at each step.)

\end{remark}

\section{Simulation Studies}

In this section, we performed two simulation studies to evaluate the performances of our suggested two-stage Bonferroni method as a high-dimensional global testing method
and  as an FWER controlling method.
\subsection{Numerical comparison for high dimensional global tests}

We performed a simulation study to compare the performance of our suggested modified two-stage Bonferroni method (See Section 5) with those of several existing global testing methods with respect to type 1 error rate and power. The methods we chose for comparison include the conventional Bonferroni test, Simes test (Simes, 1986), Higher Criterion method (Donoho and Jin, 2004, 2015), and sample-split Bonferroni test (Cox, 1975; Skol et al, 2006).

Each simulated data set is obtained by generating $m = 1000$ dependent normal random samples $N(\mu_i , \sigma^2) (i = 1, \ldots ,m)$, with a common correlation $\rho$ and a sample size $n = 15$. Among the 1,000 mean values $\mu_i$, $0$ or $m^{1-\epsilon}$ are drawn from $U(-1, 1)$ and the remaining are equal to 0, where $0 \le \epsilon \le 1$. The common variance  $\sigma^2$ is drawn from $U(0.5, 1.5)$.  For $i = 1, \ldots, m$, consider using one-sample $t$-statistic for testing individual hypothesis $H_i: \mu_i = 0$ against $K_i: \mu_i \neq 0$. We then use the aforementioned five global testing methods for testing the global hypothesis $\bigcap_{i=1}^m H_i$ against $\bigcup_{i=1}^m K_i$ at level $\alpha = 0.05$. For our suggested modified two-stage Bonferroni method, we use the sum of squares as the selection statistic for performing selection of the individual hypotheses. The selection threshold we chose is $\hat \sigma^2 \chi_n^2(1 - m^{\gamma - 1})$ in Section 5, which roughly ensures $m^{\gamma}$ of hypotheses to be selected. For the sample-split Bonferroni test, we use one-sample $t$-statistics for both selection and testing, which are respectively constructed based on the first and second half samples. The selection threshold we chose is $t_{n/2-1}(1 - m^{\gamma - 1})$ in Section 7. In addition, we always set $\gamma = 0.5$ in the simulations.

\begin{figure}
\begin{center}
\includegraphics[scale=0.7]{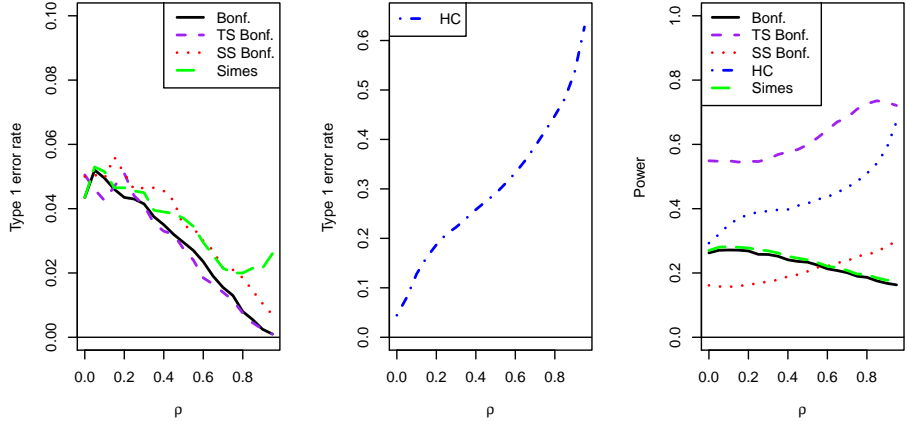}
\end{center}
\caption{Estimated type 1 error rates and powers of our suggested modified two-stage Bonferroni test (TS Bonf.) along with original Bonferroni test (Bonf.), Simes test (Simes), sample-split Bonferroni test (SS Bonf.), and Higher Criterion test (HC)
 under equal correlation $\rho$ with values from $0$ to $0.95$ and equal variance $\sigma^2 \sim U(0.5, 1.5)$. For the left and middle panels, all $\mu_i$ are equal to zero and for the right panel, $m^{1-\epsilon}$ $\mu_i$'s are drawn from $U(-1, 1)$ and the rest are equal to zero. In addition, $m = 1000$, $n = 15$ and $\alpha = 0.05$.}
\label{FWER1}
\end{figure}

The simulation is repeated for $2000$ times. The type 1 error rate and power are both estimated as the proportions of simulations where $\bigcap_{i=1}^m H_i$ is rejected when $\bigcap_{i=1}^m H_i$ is respectively true and false. In Figure 8.1 we compared the estimated type 1 error rates and powers of the aforementioned five global testing methods with respect to the common correlation. As seen from Figure 8.1, our suggested modified two-stage Bonferroni method always controls the type 1 error rate at level $\alpha$ for all values of correlation while performing best in terms of power. However, for the Higher Criterion test, it completely loses the control of type 1 error rate even when the correlation is weak; and even though for its inflated type 1 error rate, it is still less powerful than our suggested method.

In Figure 8.2 we compared the estimated power of the aforementioned five methods under independence in the cases of equal and unequal variances with respect to $\epsilon$ with values from $0.5$ to $1.0$. As seen from Figure 8.2, our suggested modified two-stage Bonferroni method performs best under equal variance in terms of power and its power improvements over the existing four methods are always pretty large for different values of $\epsilon$. Under unequal variance, our suggested modified two-stage Bonferroni method still performs well compared to the existing methods, although the power improvements become smaller when the variability of variances becomes larger.

\begin{figure}
\begin{center}
\includegraphics[scale=0.7]{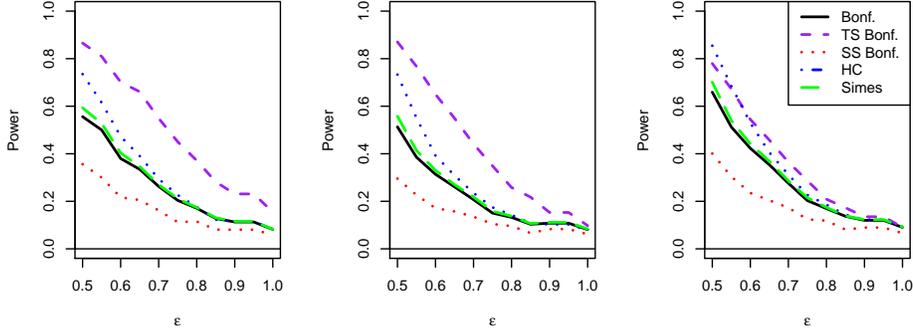}
\end{center}
\caption{Estimated powers of our suggested modified two-stage Bonferroni test (TS Bonf.) along with original Bonferroni test (Bonf.), Simes test (Simes), sample-split Bonferroni test (SS Bonf.), and Higher Criterion test (HC)
 under independence in the cases of equal variance with $\sigma_i^2 = \sigma^2 \sim U(0.5, 1.5)$ (left panel) and unequal variance with $\sigma_i^2 \sim U(0.8, 1.2)$ (middle panel) and $\sigma_i^2 \sim U(0.5, 1.5)$ (right panel). Among all these three panels, $m^{1-\epsilon}$ $\mu_i$'s are drawn from $U(-1, 1)$ with values of $\epsilon$ from $0.5$ to $1.0$  and the rest are equal to zero. In addition, $m = 1000$, $n = 15$ and $\alpha = 0.05$. }
\end{figure}

\subsection{Numerical comparison for FWER controlling procedures}

We also performed a simulation study to compare the performance of our suggested modified two-stage Bonferroni method (Section 5) with those of several existing multiple testing methods with respect to the FWER control and average power. The methods we chose for comparison include conventional Bonferroni procedure, Hochberg procedure, and sample-split Bonferroni procedure (Section 7).

Each simulated data set is obtained by generating $m = 100$
dependent normal random samples $N(\mu_i , \sigma^2) (i = 1, \ldots ,m)$, with a common correlation $\rho$ and a sample size $n = 15$. Among the 100 $\mu_i$'s, $100\pi_1$ are drawn from $U(-1, 1)$ and the remaining are equal to 0, where $\pi_1$ is the proportion of $\mu_i \neq 0$. The common variance  $\sigma^2$ is drawn from $U(0.5, 1.5)$.  For all of these four procedures, we use one-sample $t$-test statistics for testing the hypotheses $H_i: \mu_i = 0$ against $K_i: \mu_i \neq 0$. For our suggested modified two-stage Bonferroni method, we use the sum of squares as the selection statistic for performing selection of the tested hypotheses. The selection threshold we chose is $\hat \sigma^2 \chi_n^2(0.5)$, which roughly ensures about $50$ hypotheses to be selected. Here, $\hat \sigma^2$ is the average of the sample variances of the $m$ samples and $\chi_n^2(0.5)$ is the $0.5$ quantile of chi-square distribution with degrees of freedom $n$. For the sample-split Bonferroni procedure, we use one-sample $t$-statistics for performing selection of all of the $100$ hypotheses, which are constructed based on the first half sample with sample size $n_1 = 7$. The selection threshold we chose is $t_{n_1}(0.75)$, the $0.75$ quantile of $t$-distribution with degrees of freedom $n_1$, which also roughly ensures about $50$ hypotheses to be selected. For testing the selected hypotheses, we also use one-sample $t$-statistics, which are constructed based on the second half sample with sample size $n_2 = 8$.

The aforementioned four procedures are then applied to test $H_i$ against $K_i$ simultaneously for $i = 1, \ldots, 100$ at level $\alpha = 0.05$. The simulation is repeated for $2000$ times. The FWER is estimated as the proportion of simulations where at least one true null hypothesis is falsely rejected and the average power is estimated as the average proportion of rejected false null hypothesis among all false nulls across simulations. In Figure 8.3 we compared the estimated FWER and average power of these four procedures with respect to the proportion of false null hypotheses $\pi_1$ with values from $0$ to $0.5$ in the cases of $\rho = 0$ (upper panel) or $\rho = 0.5$ (bottom panel). As seen from Figure 8.3, our suggested modified two-stage Bonferroni method performs best in terms of average power while controlling the FWER at level $\alpha$, and its power improvements over the existing three methods are decreasing with the increasing proportion of false nulls.

\begin{figure}
\begin{center}
\includegraphics[scale=0.8]{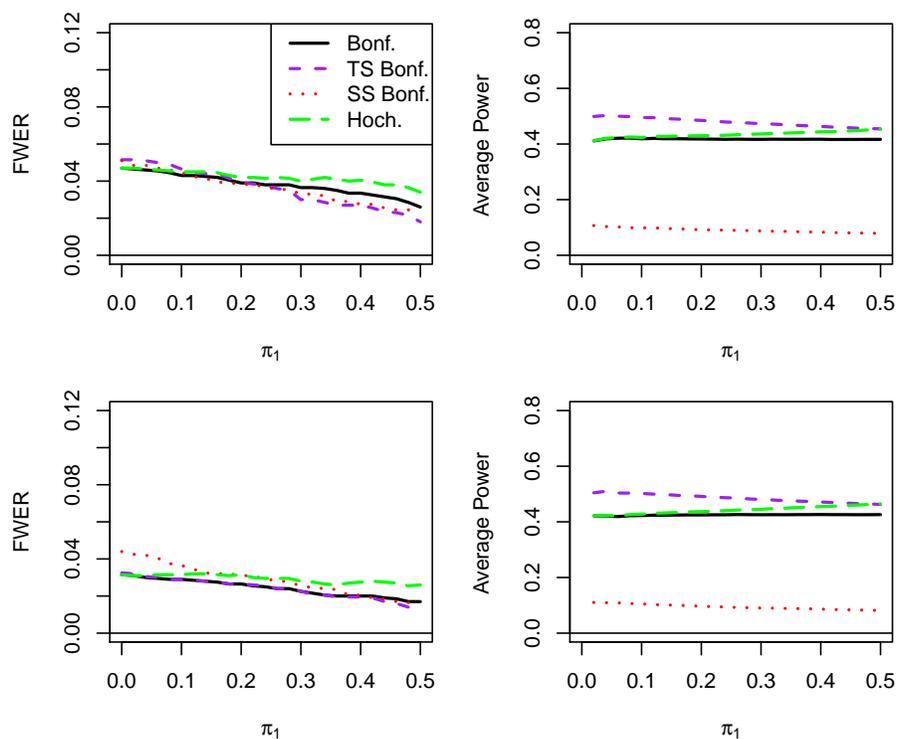}
\end{center}
\caption{Estimated FWER and powers of our suggested modified two-stage Bonferroni procedure (TS Bonf.) along with original Bonferroni procedure (Bonf.), Hochberg procedure (Hoch.), and sample-split Bonferroni procedure (SS Bonf.) under equal correlation $\rho$ with $\rho = 0$ (upper panel) or $\rho = 0.5$ (bottom panel) and equal variance $\sigma^2 \sim U(0.5, 1.5)$. For the mean values $\mu_i$,
$\pi_1 m$ $\mu_i$'s are equal to one and the rest are equal to zero. Here, the value of $\pi_1$ is from $0$ to $0.5$, $m = 100$, $n = 15$, and $\alpha = 0.05$. }
\label{FWER1}
\end{figure}

In Figure 8.4 we compared the estimated FWER and average power of these four procedures with respect to the common correlation $\rho$ with values from $0$ to $0.95$. We observe from Figure 8.4 that for different values of correlation $\rho$, our suggested modified two-stage Bonferroni method always performs best in terms of average power while controlling the FWER at level $\alpha$. In addition, we also observe that the average powers of these methods are not affected by the correlation and the estimated FWERs are basically decreasing in terms of the correlation.

\begin{figure}
\begin{center}
\includegraphics[scale=0.6]{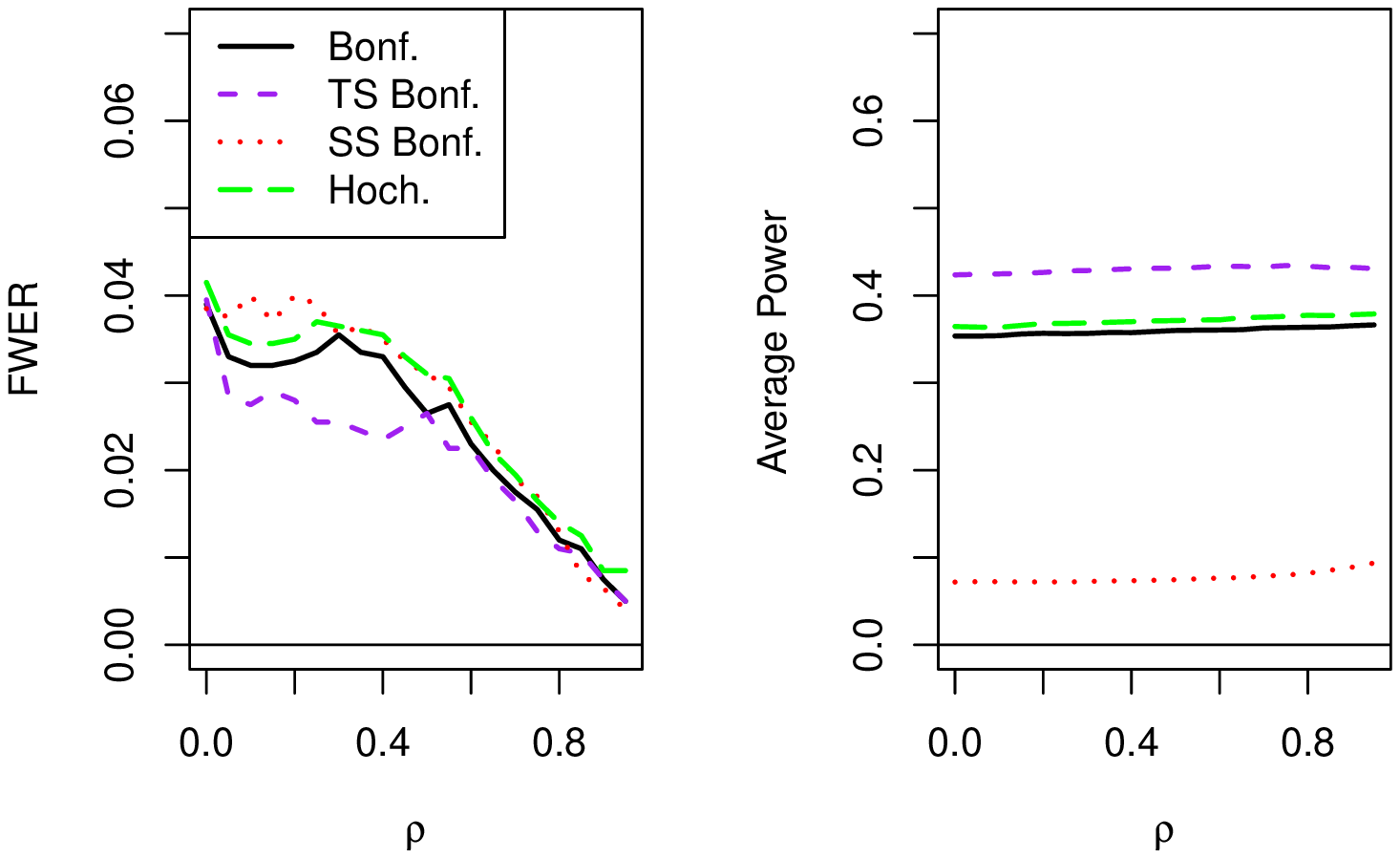}
\end{center}
\caption{Estimated FWER and powers of our suggested modified two-stage Bonferroni procedure (TS Bonf.) along with original Bonferroni procedure (Bonf.), Hochberg procedure (Hoch.), and sample-split Bonferroni procedure (SS Bonf.) under equal correlation $\rho$ with values from $0$ to $0.95$ and equal variance $\sigma^2 \sim U(0.5, 1.5)$. For the mean values $\mu_i$,
$0.2m$ $\mu_i$'s are equal to one and the rest are equal to zero. In addition, $m = 100$, $n = 15$ and $\alpha = 0.05$. }
\label{FWER1}
\end{figure}

\section{Technical Details}

\noindent{\sc Proof of Lemma  \ref{lemma:control} :}  Assume $H_i$ is true.  Then, we claim
the detection statistic $T_{n,i}$ is independent of
all the selection statistics $(S_{n,1} , \ldots , S_{n,m} )$.
For the univariate normal model with mean 0 and unknown variance, the $t$-statistic
 $T_{n,i}$ is independent of $S_{n,i}$ by Basu's theorem (because $T_{n,i}$ is ancillary and $S_{n,i}$ is a complete sufficient statistic).
 Hence, $T_{n,i}$ is independent of $S_{n,i}$, and therefore independent of $S_{n,1} , \ldots , S_{n, m}$.
  Let $I_0$ be the indices of the true null hypotheses.
Thus, the FWER is given by
\begin{equation}
FWER = P \left \{ \bigcup_{ i \in I_0}  \{ S_{n,i} > u , | T_{n,i} | > t_{n-1} ( 1- \frac{\alpha}{2 | \hat S_n | } ) \} \right \}
\end{equation}
This probability,  conditional on the selection statistics $S_{n,1} , \ldots , S_{n,m}$  is
\begin{equation}
P \left \{  \bigcup_{i \in I_0 \bigcap \hat S_n } \{  | T_{n,i} | > t_{n-1} ( 1- \frac{\alpha}{2 | \hat S_n | } ) \} \Big | S_{n,1} , \ldots , S_{n,m}  \right \}~,
\end{equation}
which by Bonferroni's inequality is bounded above by
\begin{equation}
\sum_{ i \in I_0 \bigcap \hat S_n } \alpha /  | \hat S_n | =  \frac{ | I_0 \bigcap \hat S_n |}{|\hat S_n|} \cdot \alpha \le \alpha~.
\end{equation}
Therefore, the unconditional probability is bounded above by $\alpha$, as required.~\qed

\bigskip

\noindent{\sc Proof of Theorem \ref{theorem:holm}}:
As in the proof of Lemma  \ref{lemma:control}, compute the probability of at least one false rejection conditional on the selection statistics.
Let $\hat i$ be the smallest (or first) $i$ for which $H_{r_i}$ is true and  $\tilde p_{n, r_i} \le \alpha / ( | \hat S_n | - i + 1 )$.
Such an event implies that the smallest $p$-value among the true null hypotheses which have been selected is less than or equal to
$\alpha / | \hat S_n \bigcap I_0 |$.
Indeed, the largest possible value for $\hat i$ (leading to the largest possible critical value for the first true null hypothesis tested) is given if,
out of the $| \hat S_n |$ selected hypotheses,
all
of the $| \hat S_n \bigcap I_0^c |$
false null hypotheses are rejected first in the stepdown procedure, where $I_0^c$ is the set of indices of the false null hypotheses.
This occurs when $\hat i = | \hat S_n \bigcap I_0^c | + 1$, in which case
$$\frac{\alpha}{ | \hat S_n | - \hat i + 1} = \frac{\alpha}{| \hat S_n | - ( | \hat S_n  \bigcap I_0^c | + 1) + 1}
= \frac{\alpha}{ | \hat S_n \bigcap I_0 | }~.$$
By Bonferroni, the conditional  probability is bounded above by $\alpha$
because it is the conditional probability that the  minimum of  $| \hat S_n \bigcap I_0 |$ true null $p$-values is bounded above by $\alpha / | \hat S_n \bigcap I_0 |$.   Thus, the unconditional probability of FWER is bounded above by $\alpha$.~\qed

\bigskip

Before proving Lemma \ref{lemma:select}, we will make use of the following lemmas.

\begin{lemma}\label{lemma:LM}  (Laurent and Massart, 2000). For every $n \ge 1$ and every $\beta \in (0, 1)$, we have
$$\chi_n^2(1-\beta) \le n + 2\log\left(\frac{1}{\beta}\right) + 2\sqrt{n \log\left(\frac{1}{\beta}\right)}.$$
\end{lemma}

\begin{lemma} \label{lemma:I} (Inglot, 2010). For every $n \ge 17$ and every $\beta \in [e^{-560n}, \frac{1}{17}]$, we have
$$\chi_n^2(1-\beta) \ge n + 2\log\left(\frac{1}{\beta}\right) + \frac{1}{4}\sqrt{n \log\left(\frac{1}{\beta}\right)}.$$
\end{lemma}

\noindent{\sc Proof of Lemma \ref{lemma:select}:}  To show (i), it is enough to show that $S_{n,1}$
exceeds an upper bound to $\chi_n^2 ( 1-  \beta)$ with probability tending to one.
By Lemma \ref{lemma:LM} and the specification $\beta  = m^{\gamma -1}$, we have:
$$\chi_n^2 (1- \beta )  \le  n + 2 \log ( 1/ \beta) + 2 \sqrt{ n \log (1/ \beta ) }$$
$$
= n + 2 ( 1- \gamma ) \log(m) + 2 \sqrt{ n (1- \gamma ) \log (m) }
~.$$
Now,  if $S_{n,1}$ is normalized to form $Z_n$, then by the Central Limit Theorem, it follows that
$$Z_n \equiv \frac{S_{n,1} - ( n + n \mu_1^2 )}{\sqrt{2n + 4n \mu_1^2}} \darrow N(0,1)~.$$
Thus, it suffices to show that $Z_n \ge c_n$ with probability tending to one, where
$$c_n =  \frac{ - n \mu_1^2 + 2 ( 1- \gamma ) \log(m) + 2 \sqrt{n ( 1- \gamma ) \log (m) } }{ \sqrt{ 2n + 4n \mu_1^2}}~.$$
But,
$$c_n / \sqrt{n} =  \frac{- \mu_1^2 + 2 ( 1- \gamma) \log (m)/n + 2 \sqrt{ (1- \gamma) \log (m)/n }}{ \sqrt{ 2 + 4 \mu_1^2 }}
$$
$$\to \frac{- \mu_1^2 + 2 ( 1- \gamma) d + 2 \sqrt{ (1- \gamma) d }}{ \sqrt{ 2 + 4 \mu_1^2 }} < 0~,
$$
by the assumption on $\mu_1$.  Therefore, $c_n \to - \infty$ and so $Z_n > c_n$ with probability tending to one.

To prove (ii),  we argue similarly. By Lemma \ref{lemma:I},  when $n$ is sufficiently large, we have
\begin{eqnarray}
\chi_n^2(1-\beta) & \ge & n + 2\log \left(\frac{1}{\beta} \right) + \frac{1}{4} \sqrt{n \log \left(\frac{1}{\beta} \right)} \nonumber \\
& = & n + 2(1 - \gamma) \log (m) + \frac{1}{4} \sqrt{n (1 - \gamma ) \log (m)}~.
\end{eqnarray}
Therefore, it suffices to show
$$S_{n,1} > n + 2 ( 1- \gamma ) \log (m) + \frac{1}{4} \sqrt{n ( 1- \gamma ) \log (m)}$$
with probability tending to 0.
In terms of $Z_n$, it suffices to show $Z_n \ge d_n$ with probability tending to 0, where
$$d_n =  \frac{ - n \mu_1^2 + 2 ( 1- \gamma ) \log (m) + \frac{1}{4} \sqrt{n ( 1- \gamma ) \log (m)}}
{ \sqrt{ 2n + 4n \mu_1^2}}~.$$
But
$$d_n / \sqrt{n}  \to \frac{ - \mu_1^2 + 2 ( 1- \gamma ) d + \frac{1}{4} \sqrt{(1- \gamma ) d } }
{ \sqrt{ 2 + 4 \mu_1^2 }} > 0~.$$
Hence, $d_n \to \infty$ and the result follows.~\qed

\medskip

\noindent{\sc Proof of Lemma \ref{lemma:cheb}:}  For $i = 1, \ldots, m$, let $I_i = I \{ S_{n,i}  \ge \chi_n^2(1- \beta ) \} $, where $I \{ \cdot \} $ denotes the indicator function.
Recall that  the number of selected hypotheses is  $| \hat S_n |$,  so $| \hat S_n | = \sum_{i=1}^m I_i$.
  Note that, for any true null hypothesis $H_i$,  $S_{n,i}  \sim \chi_n^2$, in which case
  $$Pr \{ S_{n,i}  \ge \chi_n^2(1-\beta ) \} = \beta~,$$ where $$\beta  = m'/m = m^{-(1-\gamma)}~.$$
  Then,
  if $H_i$ is true,
  $E(I_i) = \beta$ and $Var(I_i) = \beta (1-\beta )$.
   In fact, since the Chi-squared family of distributions  (with fixed degrees of freedom and varying noncentrality parameter) has monotone likelihood ratio, its power function is increasing in the noncentrality parameter; thus, $E( I_i ) \ge \beta$ regardless of whether or not $H_i$ is true.
 So,
 $$E ( | \hat S_n | ) \ge m \beta = m' = m^{\gamma} \to \infty~,$$
 as stated in (\ref{equation:selecta}) of  the lemma.

Now,
$$E(| \hat S_n | ) =  \sum_{i \in  I_{m,0}} E(I_i)  +
 \sum_{i \in  I_{m,1}} E(I_i)   =
 | I_{m,0} | \beta + \sum_{i \in  I_{m,1}} E(I_i)~
 $$
 So,
 \begin{equation}\label{equation:firstm}
  \beta |I_{m,0} | \le E ( | \hat S_n | ) \le
  m \beta + |I_{m,1} |  = m^{\gamma} + | I_{m,1} |~.
  \end{equation}
 Thus,
 \begin{equation}\label{equation:comb1}
  E ( | \hat S_n | / m^{\gamma} ) - 1 \le | I_{m,1} | / m^{\gamma} = O  ( m^{1- \epsilon - \gamma } ) = o (1)~,
  \end{equation}
  as long as $\epsilon + \gamma > 1$.
 Combining (\ref{equation:comb1} ) and (\ref{equation:selecta}) yields
 \begin{equation}\label{equation:comb3}
 E ( | \hat S_n | / m^{\gamma} ) \to 1~.
 \end{equation}

Using indicators again to approximate the variance of $| \hat S_n |$ yields
$$Var ( | \hat S_n | ) = \sum_{i=1}^m E (I_i ) [1 - E( I_i ) ] \le  E ( | \hat S_n | )~.$$
Therefore, making use of  (\ref{equation:comb3}).
$$Var ( | \hat S_n | / m^{\gamma} ) \le E ( | \hat S_n | ) / m^{2 \gamma} = O ( m^{- \gamma } )  \to 0~.$$
Thus, by Chebychev's inequality,
$| \hat S_n | / m^{\gamma} \Parrow 1$, yielding (\ref{equation:cheb}).  Combining (\ref{equation:cheb}) and (\ref{equation:comb3}) yields (\ref{equation:selectb}).
~\qed

\medskip

\subsection{The probability of detecting $\mu_1$}

In the second stage of the two-stage method, we need to be able to approximate
the very upper tail quantiles of the normal and  $t$ distributions.
The approximation $z_{1- \alpha /m} \approx \sqrt{2 \log (m)}$ is well-known for large $m$.
In our application, we will apply this with random $m$, and so some care must be taken to
get good lower and upper bounds to the quantile.

\begin{lemma}\label{lemma:normalq}
For any fixed $\alpha$ and any $\delta > 0$, the following inequalities hold for all large enough $m$:
\begin{equation}\label{equation:basicz}
\sqrt{ ( 1- \delta ) 2 \log (m)}  \le z_{1- \frac{\alpha}{m}}  \le \sqrt{2 \log (m)}~.
\end{equation}
\end{lemma}

\begin{remark}\rm
In fact the approximations hold uniformly for $\alpha \in [ \eta , 1- \eta ]$ for any $\eta > 0$
and for all large enough $m$.
\end{remark}

\noindent {\sc Proof of Lemma \ref{lemma:normalq}:}
If  $\phi ( \cdot )$ denotes the standard normal density and $Z \sim N(0,1)$, then the
following inequalities are well-known (see Feller (1968), Lemma 2 in Chapter VII): for any $t > 0$,
\begin{equation}
( \frac{1}{t} - \frac{1}{t^3})  \phi (t) < P \{ Z \ge  t \} \le   \frac{\phi ( t)}{ t}~.
\end{equation}
It follows from the right inequality  that
$$P \{ Z \ge \sqrt{2 \log (m)} \} \le \frac{\phi ( \sqrt{ 2 \log (m)} )}{ \sqrt{ 2 \log (m)}}
= \frac{1}{ 2  m \sqrt{ \pi  \log (m)} }
< \alpha /m$$
as soon as $\sqrt{ \log (m)} > 1/ (2 \sqrt{ \pi } \alpha )$.
Therefore, the $1- \alpha /m$ quantile of the standard normal distribution must be bounded
above by $\sqrt{2 \log (m)}$ as soon as $\sqrt{ \log (m)} > 1/ (2 \sqrt{ \pi } \alpha )$.
The first inequality is similar. ~\qed

Let $F_n$ be the cdf of student's $t$ with $n$ degrees of freedom, and $\Phi$ be the cdf of $N(0, 1)$. Consider the equation $F_n(x) = \Phi(u)$ and let $x_n(u)$ be the solution of the equation. Let
$$L_n(u) = \sqrt{n}\left (e^{\frac{u^2}{n}} - 1 \right)^{\frac{1}{2}}$$
and
$$U_n(u) = \sqrt{n}\left (e^{\frac{u^2}{n-0.5}} - 1 \right)^{\frac{1}{2}}~.$$
We will make use of the following result.

\begin{lemma}\label{lemma:FM} (Fujikoshi and Mukaihata, 1993). For all $u > 0$, we have
\begin{itemize}
  \item[(i)] $x_n(u) \ge L_n(u)~~ (n > 0);$
  \item[(ii)] $x_n(u) \le U_n(u)~~ (n > 0.5).$
\end{itemize}
\end{lemma}

As before, let $z_{1 - \alpha }$ and $t_{n-1} (1- \alpha )$ denote  the $1  - \alpha $ quantiles of $N(0, 1)$ and $t_{n-1}$, respectively.
Then
$$F_{n-1} (t_{n-1} (1 -  \alpha ) ) = \Phi( z_{1- \alpha } ) = 1 - \alpha~.$$

\begin{lemma}\label{lemma:tbound}
Fix any  $0 < \alpha < 1$ and $\delta  > 0$. Then,  for all $m$ large enough,
\begin{equation}\label{equation:boundL}
t_{n-1} ( 1- \frac{\alpha}{m} ) \ge \sqrt{n-1} \left [ \exp (  \frac{ (1- \delta ) 2 \log (m)}{n - 1} )  - 1 \right ]^{1/2}
\end{equation}
and
\begin{equation}\label{equation:boundR}
t_{n-1} ( 1- \frac{\alpha}{m} ) \le \sqrt{n-1} \left [ \exp (  \frac{2 \log (m)}{n - 1.5} )  - 1 \right ]^{1/2}~.
\end{equation}
\end{lemma}

\noindent{\sc Proof of Lemma \ref{lemma:tbound}:}   First, we show (\ref{equation:boundR}).  By Lemma \ref{lemma:FM}, we have
$$ t_{n-1}
( 1- \frac{ \alpha}{m} ) \le U_{n-1} (z_{1- \frac{\alpha }{m}} )~.$$
But since $U_{n-1} ( \cdot )$ is an increasing function, we can replace $z_{ 1-  \frac{\alpha}{m} }$
by the upper bound  $\sqrt{ 2 \log (m)}$  provided for in Lemma \ref{lemma:normalq}, at least
for all large $m$. This gives the bound on the right side of (\ref{equation:boundR}).

Similarly, for all large $m$, we have

$$ t_{n-1}
( 1- \frac{ \alpha}{m} )
 \ge L_{n-1}  ( z_{1- \frac{\alpha}{m} } )  \ge L_{n-1} ( \sqrt{ (1- \delta ) 2 \log (m)}  )~,
$$
which gives the lower bound in (\ref{equation:boundL}).
 \qed

\medskip

\noindent{\sc Proof of Lemma \ref{lemma:detect}:} To prove (i),
detection occurs when $| T_{n,1} |$ exceeds $t_{n-1} ( 1- \alpha/ 2  | \hat S_n |  )$, where
$| \hat S_n |$ is the number of selected hypotheses from the first stage.
By  Lemma \ref{lemma:cheb}, $| \hat S_n | \Parrow \infty$, and so by Lemma \ref{lemma:tbound},
$$t_{n-1} ( 1- \alpha/ 2 | \hat S_n |  ) \le \sqrt{n-1} \left [ \exp (  \frac{2 \log (2 | \hat S_n |)}{n - 1.5} )  - 1 \right ]^{1/2}
$$
with probability tending to one.
Hence,
$$P_{\mu_1} \{ |T_{n,1}|  > t_{n-1} ( 1 - \frac{ \alpha}{2 | \hat S_n |} ) \}  =  P \{  |  \frac{ \sqrt{n} ( \bar X_{n,1} - \mu_1 )}{\hat \sigma_{n,1}} + \frac{ \sqrt{n} \mu_1}{\hat \sigma_{n,1}}  |  >
t_{n-1} ( 1 - \frac{ \alpha}{2 | \hat S_n |} ) \}
$$
$$ \ge P \{|  t_{n-1} + \frac{ \sqrt{n} \mu_1}{\hat \sigma_{n,1}}  |  > \sqrt{n-1} \left [ \exp (  \frac{2 \log (2 | \hat S_n |)}{n - 1.5} )  - 1 \right ]^{1/2}
    \} + o(1)~,$$
where $t_{n-1}$ denotes a generic random variable having the $t$-distribution with $(n-1)$ degrees of freedom.
The quantity inside the probability to the left of $>$ divided by $\sqrt{n}$ tends in probability to $| \mu_1 / \sigma | = | \mu_1 |$,  i.e.,
$$| \frac{t_{n-1}}{\sqrt{n}} + \frac{\mu_1}{\hat \sigma_{n,1}} | \Parrow | \mu_1 |~.$$
But, using Lemma \ref{lemma:cheb} and Assumption A,  the quantity inside the probability to the right of  $>$ divided by $\sqrt{n}$
tends in probability to  $ \sqrt{  \exp ( 2 \gamma d) -1}$.
Hence,  by Slutsky's theorem, the probability will tend to one if $\mu_1^2 > \exp ( 2 \gamma d ) -1$.

Similarly, to prove (ii), with probability tending to one we have
$$t_{n-1} ( 1- \alpha/ 2 | \hat S_n |  ) \ge \sqrt{n-1} \left [ \exp (  \frac{ ( 1- \delta ) 2 \log ( 2 | \hat S_n |)}{n - 1} )  - 1 \right ]^{1/2}~.$$
Call the expression on the right side $\hat r_n$.  Then, the detection probability can be bounded
above as
$$P_{\mu_1} \{ |T_{n,1}|  > t_{n-1} ( 1 - \frac{ \alpha}{2 | \hat S_n |} ) \}  \le P \{ |  t_{n-1} + \frac{ \sqrt{n} \mu_1}{\hat \sigma_{n,1}} | > \hat r_n \}~.
$$
Note that the left side inside the last probability   divided by $\sqrt{n}$ tends in probability to  to $| \mu_1 / \sigma | = | \mu_1 |$, while
the right side, $\hat r_n$ divided by $\sqrt{n}$ tends in probability to $\sqrt{ \exp [ ( 1- \delta) 2 \gamma d] -1 }$.
Hence, if for some $\gamma > 0$, we have
\begin{equation}\label{equation:delta}
\mu_1^2  <  \sqrt{ \exp [ ( 1- \delta) 2 \gamma d] -1 }~,
\end{equation}
then the probability of detection tends to 0.  By continuity, if $\mu_1^2  < \sqrt{ \exp ( 2 \gamma d) -1}$, then we can choose $\delta $ small enough so that
(\ref{equation:delta}) holds, and  the result follows. ~\qed

\bigskip

\noindent{\sc Proof of Lemma \ref{lemma:cstar}}
We first argue that for any $\alpha > 0$ and $n$ sufficiently  large,
\begin{equation}\label{equation:oldlemma4}
\chi_n^2(1-\alpha) \le n + 2\log\left(\frac{1}{\alpha}\right) + c\sqrt{n \log\left(\frac{1}{\alpha}\right)},
\end{equation}
where $c$ is a given positive constant satisfying $0 < c < 2$.
Along the proof of Theorem 4.1 in Inglot (2010), to prove the above inequality, it is enough to show that
$$n\left(c \sqrt{v} - \log(1 + 2v + c \sqrt{v})\right) + 2\log\left(\frac{2}{\sqrt{n}} + \frac{2t}{\sqrt{n}} + c \sqrt{t}\right) + \log \pi \ge 0,$$
where $t = \log \left(\frac{1}{\alpha}\right)$ and $v = \frac{t}{n}$.
Then, it
 is in turn enough to show the following inequality when $n$ is sufficiently large,
\begin{equation}\label{equation:old21}
c \sqrt{v} > \log(1 + 2v + c \sqrt{v}).
\end{equation}
But for given $v$, $v > a(c)$ is equivalent to $g(c\sqrt{v}) > 2/c^2$, which in turn implies the inequality (\ref{equation:old21}).  Therefore, (\ref{equation:oldlemma4}) holds if $\frac{1}{n} \log\left(\frac{1}{\alpha}\right) > a(c)$, where $a(c)$ is defined in (\ref{equation:ac}).

Specifically, if $\alpha  = \beta_m  = m^{-(1-\gamma)}$, under assumption A, we have $v = \frac{1}{n} \log\left(\frac{1}{\alpha}\right) \to (1-\gamma)d$. Thus, for given $c \in (0, 2)$ and sufficiently large $n$, as $(1-\gamma)d > a(c)$,  (\ref{equation:oldlemma4}) holds.
 Thus, as $c \in (c^*, 2)$,  (i) holds.

 To prove (ii), the proof is similar.  When $n$ is sufficiently large, the lower bound of $\chi_n^2(1-\alpha)$ in Lemma \ref{lemma:I} can be improved as
$$\chi_n^2(1-\alpha) \ge n + 2\log\left(\frac{1}{\alpha}\right) + c\sqrt{n \log\left(\frac{1}{\alpha}\right)},$$
where $c \in (1/4, 2)$.

By using the similar arguments as in the proof of Theorem 5.2 of Inglot (2010) and wherein letting $u^* = n + 2t + c\sqrt{nt}$, to prove the above inequality, it is enough to show that
$$n\left(\log(1 + 2v + c \sqrt{v}) - c \sqrt{v}\right) - \log n - 2\log\left(\frac{2}{\sqrt{n}} + \frac{2t}{\sqrt{n}} + c \sqrt{t}\right)\ge \kappa,$$
where $\kappa = -2\log((1 - e^{-2})/2)$.

When $n$ is sufficiently large, we only need to show that
$$\log(1 + 2v + c \sqrt{v}) > c \sqrt{v},$$
which is equivalent to $v < a(c)$. Therefore, when $n$ is sufficiently large, we have
$$\chi_n^2(1-\alpha) \ge n + 2\log\left(\frac{1}{\alpha}\right) + c\sqrt{n \log\left(\frac{1}{\alpha}\right)}$$
if $\frac{1}{n} \log\left(\frac{1}{\alpha}\right) < a(c)$.

Specifically, if $\alpha = \beta_m$, by using a similar argument as above, we have
\begin{equation}
\chi_n^2(1-\beta_m) \ge n + 2\log\left(\frac{1}{\beta_m}\right) + c\sqrt{n \log\left(\frac{1}{\beta_m }\right)}
\end{equation}
for $c \in (0, c^*(\gamma, d))$.~\qed

\medskip

\begin{lemma}\label{lemma:tri}
Let $(C_1, C_2, C_3)$ have the trinomial distribution based on $n$ trials and corresponding
success probabilities $(p_1 , p_2 , p_3)$.
Then,
\begin{equation}\label{equation:tribound}
E \left ( \frac{C_1}{\max (1 , C_2 ) } \right ) \le 2 \cdot \frac{p_1}{p_2}~.
\end{equation}
\end{lemma}

\noindent{\sc Proof of Lemma \ref{lemma:tri}:}
Since $1/  \max ( 1 , C_2 ) \le 2/ ( C_2 + 1)$, it suffices to show
\begin{equation}\label{equation:tri2}
E \left ( \frac{C_1}{C_2 + 1} \right ) \le  \frac{p_2}{p_1}~.
\end{equation}
The conditional distribution of $C_2$ given $C_1$ is $c$ is binomial based on
$t = n-c$ trials and success probability $ \theta = p_2 / (1 - p_1)$.   Hence,
$$
E \left ( \frac{C_1}{C_2 + 1} | C_1 = c  \right ) =  c \sum_{j=0}^t \frac{1}{j+1} {t \choose j } \theta^j
( 1- \theta )^{t-j} = \frac{c}{(t+1) \theta } \sum_{j=0}^t { {t+1} \choose {j+1}} \theta^{j+1} (1 - \theta )^{t-j}~.
$$
The last sum is bounded above by  one because if the sum included the index $j=t+1$
the sum would be the sum of binomial probabilities based on $t+1$ trials with success parameter $\theta$.    Thus,
$$
E \left ( \frac{C_1}{C_2 + 1} | C_1 = c  \right ) \le c / [ (n-c+1) \theta ] ~ $$
and so
$$E \left ( \frac{C_1}{C_2 + 1} \right ) \le \frac{1}{\theta} E \left ( \frac{C_1}{n-C_1+1} \right )
= \frac{1}{\theta} \sum_{j=0}^n \frac{j}{n-j+1} {n \choose j} p_1^j ( 1- p_1 )^{n-j}$$
$$
= \frac{p_1}{\theta (1- p_1 )} \sum_{i=0}^{n-1}  { n \choose {i}} p_1^{i} (1-  p_1 )^{n - i}
\le \frac{p_1}{\theta (1- p_1 )} = \frac{ p_1}{ p_2}~.~~\qed$$

\bigskip

\noindent{\sc Proof of Theorem \ref{theorem:esigma}:} Without loss of generality, assume $\sigma = 1$.
Also note that the FWER is maximized when all null hypotheses are true.
Indeed, the number of hypotheses selected is an increasing function of $| \mu_i |$, where $\mu_i$
is the mean of the $i$th sample (since the non-central Chi-squared distribution has monotone
likelihood ratio in the non-centrality parameter).  But increasing the number of selections
only makes the FWER smaller since  (stochastically) more hypotheses are tested
at the second stage than just the true nulls.  Hence, we now assume all hypotheses are null.

For any $\tau_n \to \infty$,  the event $E_n$
defined by
\begin{equation}\label{equation:En}
E_n = \left  \{ 1 - \frac{\tau_n}{\sqrt{mn}} \le \hat \sigma^2 \le 1 + \frac{\tau_n}{\sqrt{mn}} \right \}~
\end{equation}
has probability tending to one.
Let $\delta_n = \tau_n / \sqrt{mn}$.
For any $u$, let
$$I_n (u ) = \{ i:~ S_{n,i} > u \}~,$$
be the selection set when it is known $\sigma = 1$;  in particular, we will
always take $u = \chi^2 ( 1- m^{\gamma -1 } )$.
Then, with probability tending to one,
\begin{equation}\label{equation:subs1}
I_n ( u + \delta_n u ) \subseteq \hat I_n ( u) \subseteq I_n ( u - \delta_n u )
\end{equation}
and correspondingly the numbers of elements in these index sets satisfy
\begin{equation}\label{equation:subs2}
| I_n ( u + \delta_n u  ) | \le |  \hat I_n ( u)| \le |  I_n ( u - \delta_n u )|~.
\end{equation}
Then, using (\ref{equation:subs1}) and (\ref{equation:subs2}),
\begin{equation}\label{equation:FWER2}
FWER = P \left  \{ \bigcup_{i \in \hat I_n (u) }  \{ |T_{n,i} | > t_{n-1, 1 - \frac{\alpha}{2 \max (1, | \hat I_n (u) |)}}  \} \right \}
\end{equation}
$$ \le P \left  \{ \bigcup_{i \in  I_n (u - \delta_n u) }  \{ |T_{n,i} | > t_{n-1, 1 - \frac{\alpha}{2 \max (1,  |  I_n (u + \delta_n u ) |)}}  \} \right \}
+ P ( E_n^c )~.$$
The point is that, conditional on all the $S_{n,i}$, the sets $I_n ( \cdot )$
are determined, and the $t$-statistics then remain conditionally independent (but not so
if we condition on $\hat I_n ( u )$).
Hence, by the Bonferroni inequality,  the last probability, conditional on the $S_{n,i}$,
is bounded above by $\alpha | I_n ( u - \delta_n u ) |  / \max (1, | I_n ( u + \delta_n u ) | $.
Hence, to complete the argument, we must show
\begin{equation}\label{equation:show}
E \frac{ | I_n ( u - \delta_n u ) | }{ \max (1, | I_n ( u + \delta_n u ) |)} \to 1~.
\end{equation}
Let $C_1$ be the number of $S_{n,i}$ in $(u- \delta_n u , u + \delta_n u )$ and $C_2$
be the number $\ge u + \delta_n u$.
Then, (\ref{equation:show}) reduces to showing
$$E  \left ( \frac{ C_1 + C_2 }{\max (1, C_2 )} \right ) \to 1$$
or equivalently
$$E  \left ( \frac{ C_1  }{\max (1, C_2 )} \right ) \to 0~.$$
By Lemma \ref{lemma:tri}, this last expression is bounded above by $2 p_1 / p_2$,
and so we must show $p_1 / p_2 \to 0$, where
\begin{equation}\label{equation:ratio}
\frac{p_1}{p_2} =  \frac{  P \{ S_{n,i} \in ( u - \delta_n u , u + \delta_n u  ) \} }{ P \{ S_{n,i} > u + \delta_n u \}}~.
\end{equation}
But,  the denominator in (\ref{equation:ratio}) satisfies
$$P \{ S_{n,i} > u + \delta_n u \} \ge P \{ S_{n,i} > u \} - P \{ S_{n,i} \in ( u - \delta_n u , u + \delta_n u ) \}$$
and so it suffices to show
\begin{equation}\label{equation:nratio}
 \frac{ P \{ S_{n,i} \in ( u - \delta_n u , u + \delta_n u  ) \} }{
P \{ S_{n,i} > u  \}} \to 0~.
\end{equation}
The denominator in (\ref{equation:ratio}) is, by construction, $\beta = m^{\gamma -1}$.
The numerator involves an integration over $f_n ( \cdot )$,  the Chi-squared density with $n$ degrees of freedom.
The mode of $f_n ( \cdot )$ is $n -2$.
So, the integral can crudely be bounded above by $f_n ( n-2 )$,  the density at the mode, multiplied  by the length
of the interval ($2 \delta_n u $).
But,
$$f_n ( n-2 ) =  \frac{1}{ 2^{ \frac{n}{2}} \Gamma ( \frac{n}{2} ) } (n-2)^{\frac{n}{2}-1} e^{ - \frac{1}{2} (n-2) } ~,$$
which by Stirling's formula is easily checked to be of order $n^{-1/2}$.
Hence,  the left side  of (\ref{equation:ratio}) is bounded above by
$$\frac{2 \delta_n u \cdot \frac{1}{\sqrt{n}}
 }{m^{\gamma-1}}~.$$
Recalling that $\delta_n = \tau_n / \sqrt{nm}$ and $u = O(n)$ shows the last expression
is of order
$\tau_n  m^{ \frac{1}{2}  - \gamma } $. For $\gamma > 1/2$ and $\tau_n \to \infty$
slowly enough, this last expression tends to 0 as required.

 For $d > 0$, one can improve the argument as follows.
Note that the Chi-squared density is decreasing to the right of its mode.
Rather than using $f_n ( n-2)$, one can use $f_n (x)$ with $x$ corresponding
to (or approximating)  the point in the interval $u \pm \delta_n u$ closest to $n-2$, i.e., $u - \delta_n u$.
Note that
\begin{equation}\label{equation:bigeps}
u/n \to  1 + 2 (1- \gamma ) d + c^* ( \gamma ,d) \sqrt{ (1- \gamma ) d}  >  1+ \epsilon
\end{equation}
for some $\epsilon > 0$; thus,  $u - \delta_n u \ge (1+ \epsilon ) n$ for all large $n$.
Thus, we can bound the numerator in (\ref{equation:nratio}) by the length of the interval, $2 \delta_n u$ multiple by the density at the value $n ( 1 + \epsilon )$  of the Chi-squared distribution with $n$
degrees of freedom.
But, the  Chi-squared density evaluated at $n ( 1 + \epsilon  )$ is equal to
$$\frac{1}{ 2^{ \frac{n}{2}} \Gamma ( \frac{n}{2} ) } [n ( 1 + \epsilon ) ]^{\frac{n}{2}-1} e^{ - \frac{1}{2} n ( 1 + \epsilon )} $$
which by Stirlings formula is of order
$$\frac{ e^{-n \epsilon  /2} (1+ \epsilon )^{n/2}}{\sqrt{n}}~.$$

Hence, the expression  (\ref{equation:nratio})  is bounded above by  order
$$\frac{2 \delta_n u \cdot \frac{1}{\sqrt{n}} e^{-n \epsilon /2 } (1+ \epsilon )^{n/2}  }{m^{\gamma-1}}~.$$
Recalling that $\delta_n = \tau_n / \sqrt{nm}$ and $u = O(n)$ shows the last expression
is of order
\begin{equation}\label{equation:tired}
\tau_n  m^{ \frac{1}{2}  - \gamma } e^{-n \epsilon /2 } (1+ \epsilon )^{n/2}~.
\end{equation}
Now, even for $\gamma = 1/2$, this last expression (\ref{equation:tired})  tends to 0 for $\tau_n \to \infty$
sufficiently slowly, since  $e^{-n \epsilon /2 } (1+ \epsilon )^{n/2} \to 0$.

Note (\ref{equation:tired}) is equal to
$$\tau_n \exp [ ( \frac{1}{2} - \gamma ) \log (m) - \frac{n \epsilon}{2} + \frac{n}{2} \log ( 1 + \epsilon ) ]
$$
$$ = \tau_n  \exp \{ n  [ ( \frac{1}{2} - \gamma ) d- \frac{ \epsilon}{2} + \frac{1}{2} \log ( 1 + \epsilon ) ]  + o(1) \}
$$
Hence, this last expression will tend to 0 (with $\tau_n \to \infty$ sufficiently slowly) if
\begin{equation}\label{equation:abc}
( \frac{1}{2} - \gamma ) d- \frac{ \epsilon}{2} + \frac{1}{2} \log ( 1 + \epsilon ) <  0~.
\end{equation}
But by (\ref{equation:bigeps}), we can take any $\epsilon$ satisfying
\begin{equation}\label{equation:pester}
\epsilon <   2 (1- \gamma ) d + c^* ( \gamma ,d) \sqrt{ (1- \gamma ) d} ~.
\end{equation}
Therefore, if we let $\epsilon^*$ be the right side of (\ref{equation:pester}), then the result
will follow for any $\gamma$ satisfying (\ref{equation:abc}) with $\epsilon $ replaced
by $\epsilon^*$, as claimed.~\qed

\bigskip

\noindent{\sc Proof of  Theorem \ref{theorem:dependence1}}:  For every $0 < \varepsilon < \pi_0$, let $E_{n, 1}$ denote the event $\{| \hat S_{n,0} | \ge (\pi_0 - \varepsilon)m^\gamma\}$.  Under assumption B2, we have
\begin{equation}\label{equation:asymp}
P(E_{n, 1}^c) \rightarrow 0 ~~~ \text{as  } m \rightarrow \infty.
\end{equation}
Thus, the FWER is given by
\begin{eqnarray}
FWER & = & P \left \{ \bigcup_{ i \in I_{m, 0}}  \{ S_{n,i} > u , | T_{n,i} | > t_{n-1} ( 1- \frac{\alpha}{2 |\hat S_n| } ) \} \right \} \nonumber \\
& \le & P \left \{ \bigcup_{ i \in I_{m, 0}}  \{ S_{n,i} > u , | T_{n,i} | > t_{n-1} ( 1- \frac{\alpha}{2 |\hat S_n| } ) \} \bigcap E_{n, 1} \right \}  + P \left \{E_{n, 1}^c \right \} \nonumber \\
& \le & \sum_{ i \in I_{m, 0}} P \left \{ S_{n,i} > u \right \} P \left \{| T_{n,i} | > t_{n-1} ( 1- \frac{\alpha}{2 (\pi_0 - \varepsilon)m^\gamma } )  \right \}  + P \left \{E_{n, 1}^c \right \} \nonumber \\
& = & \frac{\alpha}{(\pi_0 - \varepsilon)m^\gamma } E\{| \hat S_{n,0} |\}   + P \left \{E_{n, 1}^c \right \} \nonumber \\
& \rightarrow & \frac{\pi_0\alpha}{\pi_0 - \varepsilon}~~~~     \text{ as } m \rightarrow \infty, \nonumber \\
& \rightarrow &  \alpha~~~~     \text{ as } \varepsilon \rightarrow 0. \nonumber
\end{eqnarray}
Here, the second inequality follows from independence of $S_{n,i}$ and $T_{n,i}$ when $H_i$ is true, and the second last expression follows from (\ref{equation:assumB1}) and (\ref{equation:asymp}). ~\qed

\bigskip

\noindent{\sc Proof of  Theorem \ref{theorem:dependence2}}: Let $\delta_n = \frac{\tau_n}{\sqrt{mn}}$ for some $\tau_n \rightarrow \infty$ slowly such that under assumption B3, the event $E_{n, 1}$ defined by $E_{n, 1} = \{\hat \sigma^2 \ge (1 - \delta_n) \sigma^2\}$ has probability tending to one. For any $0 < \varepsilon < \pi_0$,
let $E_{n, 2}$ denote the event $\{|\hat S_{n,0}(1 - \delta_n)| \ge (\pi_0 - \varepsilon)m^\gamma\}$. Under assumption B4, the event $E_{n, 2}$ has also probability tending to one. Thus,
\begin{equation} \label{equation:limit}
\lim_{m \rightarrow \infty} P(E_{n, 1}^c) = 0~ \text{ and } ~\lim_{m \rightarrow \infty} P(E_{n, 2}^c) = 0~.
\end{equation}
We still use $\hat S_n$ to denote the indices of selected hypotheses, i.e., indices $i$ such that $S_{n,i}  > \hat \sigma^2 u$. Thus, the FWER is given by
\begin{eqnarray}
& & FWER = P \left \{ \bigcup_{ i \in I_{m, 0}}  \{ S_{n,i} > \hat \sigma^2 u , | T_{n,i} | > t_{n-1} ( 1- \frac{\alpha}{2 |\hat S_n| } ) \} \right \} \nonumber \\
& \le & P \left \{ \bigcup_{ i \in I_{m, 0}}  \{ S_{n,i} > \hat \sigma^2 u , | T_{n,i} | > t_{n-1} ( 1- \frac{\alpha}{2 |\hat S_n| } ) \} \bigcap E_{n, 1} \bigcap E_{n, 2} \right \}  \nonumber \\
& & \quad + ~P \left \{E_{n, 1}^c \bigcup E_{n, 2}^c \right \} \nonumber \\
& \le & \sum_{ i \in I_{m, 0}} P \left \{ S_{n,i} >  (1 - \delta_n) \sigma^2u \right \} P \left \{| T_{n,i} | > t_{n-1} ( 1- \frac{\alpha}{2 (\pi_0 - \varepsilon)m^\gamma } )  \right \}  \nonumber \\
& & \quad + ~P \left \{E_{n, 1}^c \right \} + P \left \{E_{n, 2}^c \right \} \nonumber \\
& = & \frac{\alpha}{(\pi_0 - \varepsilon)m^\gamma } E\{|\hat S_{n,0}(1 - \delta_n)|\}   + P \left \{E_{n, 1}^c \right \} + P \left \{E_{n, 2}^c \right \} \nonumber \\
& \rightarrow & \frac{\pi_0\alpha}{\pi_0 - \varepsilon}~~~~     \text{ as } m \rightarrow \infty, \nonumber \\
& \rightarrow &  \alpha~~~~     \text{ as } \varepsilon \rightarrow 0. \nonumber
\end{eqnarray}
Here, the second inequality follows from independence of $S_{n,i}$ and $T_{n,i}$ under $H_i$ and the Bonferroni inequality, and the second last expression follows from (\ref{equation:limit}), assumption B1, and the proof of Theorem 5.1, in which it has been shown that
$$\frac{P \left \{ S_{n,i} >  (1 - \delta_n) \sigma^2u \right \}}{P \left \{ S_{n,i} >  \sigma^2u \right \}}
\rightarrow 1 \text{ as } m \rightarrow \infty~, $$
which in turn implies
$$\frac{E\{|\hat S_{n,0}(1 - \delta_n)|\} }{m^\gamma}
\rightarrow \pi_0 \text{ as } m \rightarrow \infty~. ~\qed$$

\bigskip

\noindent{\sc Proof of  Theorem \ref{theorem:bonpower}}:  The rejection probability is
\begin{equation}\label{equation:limi}
P_{\mu_1} \{  |T_{n,i} | > t_{n-1} ( 1 - \frac{\alpha}{2 m} ) \} =  P \{ | t_{n-1} + \frac{\sqrt{n} \mu_1}{  \hat \sigma_{n,1}} | >   t_{n-1} ( 1 - \frac{\alpha}{2 m} ) \} ~,
\end{equation}
where $t_{n-1}$ denotes a generic random variable having the $t$-distribution with $n-1$ degrees of freedom.
But,
$$\frac{| t_{n-1} + \frac{\sqrt{n} \mu_1}{  \hat \sigma_{n,1}} |}{\sqrt{n}} \Parrow | \mu_1 |~.
$$
Moreover, by Lemma \ref{lemma:tbound},
$$\frac{ t_{n-1} ( 1 - \frac{\alpha}{2 m} )}{\sqrt{n}} \to \left [ e^{2d} -1 \right  ]^{1/2}~.$$
Hence, the limit  of the rejection probability in (\ref{equation:limi}) equals one or zero according to whether or not $\mu_1^2$ exceeds $e^{2d} -1 $.~\qed

\bigskip

\noindent{\sc Proof of  Theorem \ref{theorem:split}}: We first show that $H_i$ is selected with probability 1 (or 0) if
$\mu_i^2$ exceeds (or is less than) $\exp ( \frac{ 2 ( 1- \gamma ) d}{r} ) - 1$.
This is the probability
$$P_{\mu_i}  \{ | T_{n,i}^{(1)} >  t_{n_1 -1} ( 1- m^{\gamma -1}/2 ) \} =
$$
$$P \{ | t_{n_1 -1} +  \frac{ \sqrt{n_1} \mu_i}{\hat \sigma_{n,i}^{(1)}} |
>
t_{n_1 -1} ( 1- m^{\gamma -1}/2 ) \}~,$$
where $t_{n_1 -1}$ denotes a random variable having the $t$-distribution with $n_1 -1$ degrees of freedom, and $\hat \sigma_{n,i}^{(1)}$
is the sample standard deviation for the $i$th component based on the first $n_1$ observations.
But,
$$\frac{ | t_{n_1 -1} +  \frac{ \sqrt{n_1} \mu_i}{\hat \sigma_{n,i}^{(1)}} | }{\sqrt{n_1}} \Parrow \mu_1$$
and, by Lemma \ref{lemma:tbound},
$$\frac{t_{n_1 -1} ( 1- m^{\gamma -1}/2 )}{\sqrt{n_1}} \to \exp ( \frac{ 2 ( 1- \gamma ) d}{r} ) - 1~,$$
and the first claim follows.

The detection analysis is the same as for Lemma \ref{lemma:detect}, except that the number of selections $| \hat S_n |$ is obtained differently.
All that is needed is that $| \hat S_n | / m^{\gamma} \Parrow 1$.  But the identical argument used to show this in Lemma \ref{lemma:cheb} applies
as well.  Thus, using the same argument in Lemma \ref{lemma:detect}, but with $n$ replaced by $n_2 \approx (1-r)n$ gives that
$H_i$ is detected or not according to as whether $\mu_i^2$ is greater or less than
$\exp ( \frac{2 \gamma d}{1-r}) -1$.  Combining this result with the first claim completes the proof. ~\qed

\vskip 10pt

\section*{Acknowledgements}
The research of the first author was supported in part
by NSF Grant DMS-1309162 and the
research of the second author was supported in part
by NSF Grant DMS-1307973.  This work began during the first author's sabbatical stay at Stanford University, and W.G. is thankful to Stanford for hosting him.

\end{document}